\documentclass[12pt]{article}
\usepackage{amsmath,amssymb,epsfig}
\makeatletter \@addtoreset{equation}{section} \makeatother
\renewcommand{\theequation}{\thesection.\arabic{equation}}
\newcommand{\ba}{\begin{array}}
\newcommand{\ea}{\end{array}}
\newcommand{\beq}{\begin{equation}}
\newcommand{\eeq}{\end{equation}}
\newcommand{\bea}{\begin{eqnarray}}
\newcommand{\eea}{\end{eqnarray}}

\def\bce{\begin{center}}
\def\ece{\end{center}}

\def\nonu{\nonumber}

\def\be{\beta}

\def\eps6{{\displaystyle \mathop{\epsilon}^{6}}{}}

\def\nab6{{\displaystyle \mathop{\nabla}^{6}}{}}

\def\0{{\sst{(0)}}}
\def\1{{\sst{(1)}}}
\def\2{{\sst{(2)}}}
\def\3{{\sst{(3)}}}
\def\4{{\sst{(4)}}}
\def\5{{\sst{(5)}}}
\def\6{{\sst{(6)}}}
\def\7{{\sst{(7)}}}
\def\8{{\sst{(8)}}}

\def\ba{\begin{array}}
\def\ea{\end{array}}
\def\beq{\begin{equation}}
\def\eeq{\end{equation}}
\def\be{\begin{equation}}
\def\ee{\end{equation}}

\def\eps{\epsilon}

\def\ba{\begin{array}}
\def\ea{\end{array}}
\def\beq{\begin{equation}}
\def\eeq{\end{equation}}
\def\be{\begin{equation}}
\def\ee{\end{equation}}

\def\eps{\epsilon}

\newcommand{\bean}{\begin{eqnarray*}}
\newcommand{\eean}{\end{eqnarray*}}

\begin{document}
\thispagestyle{empty} \addtocounter{page}{-1}
\begin{flushright}
\end{flushright}

\vspace*{1.3cm}

\centerline{ \Large \bf  More Meta-Stable Brane Configurations without
D6-Branes}
\vspace*{1.5cm}
\centerline{{\bf Changhyun Ahn} 
} 
\vspace*{1.0cm} 
\centerline{\it 
Department of Physics, Kyungpook National University, Taegu
702-701, Korea} 
\vspace*{0.8cm} 
\centerline{\tt ahn@knu.ac.kr} 
\vskip2cm

\centerline{\bf Abstract}
\vspace*{0.5cm}

We describe the intersecting brane configurations, consisting of 
NS-branes, D4-branes(and anti-D4-branes), in type IIA string
theory corresponding to the meta-stable nonsupersymmetric vacua of
${\cal N}=1$ $SU(N_c) \times SU(N_c') \times SU(N_c'')$ gauge theory
with bifundamentals. By adding the orientifold 4-plane to these brane 
configurations, we also discuss the meta-stable brane configurations
for other gauge theory with bifundamentals.
Furthermore, we study the intersecting brane configurations 
corresponding to the nonsupersymmetric 
meta-stable vacua of other gauge theory with
bifundamentals, by adding the orientifold 6-plane.

\baselineskip=18pt
\newpage
\renewcommand{\theequation}
{\arabic{section}\mbox{.}\arabic{equation}}

\section{Introduction}

In the standard type IIA brane configuration, the quark masses correspond to 
the relative displacement of the D6-branes(0123789) and
D4-branes(01236) 
along the 45 directions geometrically. Then the eigenvalues of quark mass matrix
correspond to the positions of D6-branes in 45 directions. See
the review paper \cite{GK98} for the gauge theory and the brane dynamics.
The Seiberg duality in the classical brane picture 
can be accomplished by exchanging the locations of the NS5-brane(012345) and 
NS5'-brane(012389) along $x^6$ direction each other. 

The geometric misalignment of D4-branes connecting both NS5'-brane  and
D6-branes in the magnetic brane configuration 
can be interpreted as a nontrivial F-term condition in the
gauge theory with massive flavors. 
Then the F-term equations can be partially cancelled by both recombination of 
flavor-D4-branes  with the color-D4-branes  
and then movement of those D4-branes into the 45 directions.
This phenomenon in magnetic brane configuration 
corresponds to the fact that some entries in the magnetic dual
quarks
acquire nonzero vacuum expectation values to minimize the F-term in
the dual gauge theory side. 
Moreover, the remaining flavor-D4-branes that do not move to 45
directions, connecting to 
NS5'-brane, can move along 89 directions freely since
D6-branes and NS5'-brane are parallel
and this geometric freedom of meson field corresponds to the
classical pseudomoduli space of nonsupersymmetric vacua of the gauge theory. 

On the other hand, 
it is known that the NS-brane configuration in type IIA string theory,
where 
there exist only two types of NS5-brane and 
NS5'-brane, preserves ${\cal N}=2$
supersymmetry in four dimensions \cite{GK98}.
The geometry \cite{CHS} of the coincident NS5-branes 
is characterized by 
the metric, the dilaton, and the field strength and is useful to
construct the DBI action for D4-branes. 
In order to break the supersymmetry, one adds 
D4-branes and anti-D4-branes \cite{GK}. 
By adding D4-branes suspending between the
NS5-brane and the NS5'-brane, and 
anti-D4-branes($\overline{D4}$-branes) suspending between the
NS5-brane and the other NS5'-brane, the
supersymmetry of this system is broken \cite{GK}. 
The low energy dynamics can be described by the gauge theory on the
D4-branes.
The brane configuration corresponding to the electric theory 
with vanishing mass for the bifundamentals consists of 
the left NS5'-brane, the middle NS5-brane and the right NS5'-brane and
two sets of D4-branes suspended between two NS5'-branes.
The gauge group is a product of two unitary gauge groups and there exist 
bifundamentals. For the nonvanishing mass for these bifundamentals, 
the relative displacement between the two NS5'-branes along the 45
directions occurs. By taking the Seiberg dual for one of 
two gauge group factors with nonvanishing masses for the bifundamentals,
the magnetic dual theory has a cubic superpotential between the dual
quarks and a meson which is nothing but a quadratic term of 
bifundamentals in an electric theory. 
Also the linear term in a meson appears in this
magnetic superpotential. 
Then the F-term equation for this meson field leads to the
supersymmetry breaking. 
One finds that supersymmetry is broken classically but is restored 
quantum mechanically. It turns out the classical nonsupersymmetric 
vacuum becomes long-lived state \cite{GK}.

As the distance between the two NS5'-branes along 
the 45 directions 
becomes zero, this brane configuration with
D4- and $\overline{D4}$-branes can decay and the geometric 
misalignment between 
flavor-D4-branes arises, as before.  
Due to the presence of NS5-brane in this system, there 
exists an attractive force between the tilted D4-branes and NS5-brane.
The explicit and careful 
computation of DBI action for these D4-branes in the
background created by NS5-brane has been done by the work of \cite{GK} 
and this effect of the gravitational attraction 
leads to a curve for tilted D4-branes rather than a
straight line. 
Then for small displacement of two NS5'-branes,
the ground state is given by this ``curved'' brane configuration. 
As this displacement between two NS5'-branes is increased, the ground state 
brane configuration is given by ``straight'' brane configuration.
The meta-stable vacua of \cite{ISS} arise in some
region of parameter space. In this description, 
the dual quarks are represented by the
bifundamentals of product gauge group and the mass term is encoded by 
the relative displacement of two NS5'-branes 
in 45 directions, as we mentioned before.
Note that there exist no D6-branes in this brane configuration 
\footnote{A replacement of D6-branes with NS5'-brane corresponds to
  the gauging of the flavor group(global symmetry) 
of the gauge theory realized 
on the D4-branes
and this replacement 
might be useful to construct the phenomenological model
building.}. When one of the NS5'-branes goes to infinity along the
$x^6$ direction, 
then the corresponding gauge group becomes a global symmetry 
and the theory leads to a standard ${\cal N}=1$ SQCD with fundamentals.   
In other regions, a generalization of \cite{ISS} showing 
very similar qualitative phenomena in 
classical string theory occurs \cite{GK}.

The focus on the new meta-stable brane configurations 
by adding an orientifold 4-plane and an orientifold 6-plane to 
the above brane configuration studied by \cite{GK}, along the line of 
\cite{OO,FGU,BGHSS,Ahn06}, was given in \cite{Ahn07-5}.
When the former was added, no extra NS-branes or D-branes were needed.
However, when the latter was added, the extra NS-branes or D-branes 
into the above
brane configuration were needed in order to have a product gauge group. 

In this paper, we continue to find out new meta-stable brane
configurations which contain four NS-branes or six NS-branes, along
the line of \cite{GK,Ahn07-5}, by starting from the known or new 
supersymmetric
brane configurations in type IIA string theory.
Compared with the previous approaches given by \cite{OO,FGU,BGHSS,Ahn06},
the superpotential in the magnetic theory has very simple form because
there are no D6-branes in the brane configurations 
and this fact allows us to analyze the meta-stable vacua easily using the F-term
equations and one loop effective potential. 
But the number of NS-branes is increased for a 
given gauge theory with matters since the role of D6-branes is
replaced by NS5'-brane. 
Some of the meta-stable brane configurations lead to the known
meta-stable brane configuration corresponding to the gauge theory with
less gauge group factors in the literature, by replacing the
NS5'-brane with coincident D6-branes. 
Basically, the gauge group will be a triple product gauge group for 
four NS-branes with D4-branes.
When we add an orientifold 4-plane into this brane configuration, 
the gauge group will be a triple product between symplectic gauge
group and an orthogonal gauge group, alternatively, 
depending on the orientifold
4-plane charge. When we add an orientifold 6-plane into 
six NS-branes with D4-branes,
one of the gauge group factor will be a symplectic or an orthogonal
gauge group depending on the orientifold 6-plane charge and the other
gauge group factor will be unitary.

In section 2, we describe the type IIA brane configuration 
corresponding
to the electric theory based on the ${\cal N}=1$ $SU(N_c) \times
SU(N_c') \times SU(N_c'')$ 
gauge theory 
with the bifundamentals and deform this theory by adding the mass term
for the bifundamental. 
We construct the three different 
dual magnetic theories by taking the Seiberg
dual for each gauge group factor. 
Then we consider the nonsupersymmetric meta-stable
minimum  and present 
the corresponding intersecting brane configurations of type IIA string
theory.

In section 3, we discuss the type IIA brane configuration,
by adding
the oreintifold 4-plane to the brane onfiguration in section 2, corresponding
to the electric theory based on the ${\cal N}=1$ $Sp(N_c) \times
SO(2N_c') \times Sp(N_c'')$ 
gauge theory 
with  matters and deform this theory by adding the mass term
for the bifundamental. 
Then we construct the  corresponding 
dual magnetic theories  by taking the Seiberg
dual for each gauge group factor. 
We consider the nonsupersymmetric meta-stable
minimum  and present 
the corresponding intersecting brane configurations of type IIA string
theory. We also comment on the case of 
 ${\cal N}=1$ $SO(2N_c) \times
Sp(N_c') \times SO(2N_c'')$ 
gauge theory 
with  matters.

In section 4, we discuss the type IIA brane configuration corresponding
to the electric theory based on the ${\cal N}=1$ $Sp(N_c) \times
SU(N_c') \times SU(N_c'')$ 
gauge theory 
with  matters and deform this theory by adding the mass term
for the bifundamental. 
Then we construct the  two different 
dual magnetic theories  by taking the Seiberg
dual for each unitary gauge group factor. 
We consider the nonsupersymmetric meta-stable
minimum  and present 
the corresponding intersecting brane configurations of type IIA string
theory. Moreover,
we also discuss the meta-stable brane configurations corresponding
to the electric theory based on the ${\cal N}=1$ $SO(N_c) \times
SU(N_c') \times SU(N_c'')$ 
gauge theory by changing the orientifold 6-plane charge. 

In section 5,  we make some comments for the future directions.

\section{Meta-stable brane configurations with four NS-branes}

\subsection{Electric theory}

The type IIA brane configuration \cite{BH,AT97} corresponding to 
${\cal N}=1$ supersymmetric gauge theory with
gauge group
\bea
SU(N_c) \times SU(N_c') \times SU(N_c'')
\label{gaugegroup}
\eea
and with a field $F$ charged under
$({\bf N_c, \overline{N_c'}})$, a field $G$ charged under
$({\bf N_c', \overline{N_c''}})$, and their conjugates 
$\widetilde{F}$ and $\widetilde{G}$ 
can be described by 
the left $NS5_L'$-brane(012389), the left 
$NS5_L$-brane(012345),
the right $NS5_R'$-brane(012389), the right 
$NS5_R$-brane(012345),
 $N_c$-, $N_c'$-  and $N_c''$-color D4-branes(01236). 
The fields $F$ and $\widetilde{F}$  correspond to 4-4 strings connecting 
the $N_c$-color D4-branes with $N_c'$-color D4-branes while
the fields $G$ and $\widetilde{G}$  correspond to 4-4 strings connecting 
the $N_c'$-color D4-branes with $N_c''$-color D4-branes.

The left $NS5_L$-brane is located at $x^6=0$ and let us denote the $x^6$ 
coordinates for the $NS5_L'$-brane, the $NS5_R'$-brane and the $NS5_R$-brane 
by $x^6=-y_1, y_2, y_2+y_3$
respectively. 
The $N_c$ D4-branes 
are suspended between the 
$NS5_L'$-brane and the $NS5_L$-brane, 
the $N_c'$ D4-branes 
are suspended between the 
$NS5_L$-brane and the $NS5_R'$-brane, and 
the $N_c''$ D4-branes  
are suspended between the $NS5_R'$-brane and the $NS5_R$-brane.
We draw this brane configuration in Figure 1A for the vanishing mass
case
\footnote{There are similar brane configurations in the context of
  quiver gauge theory \cite{ABFK,KOO}.}. 

\begin{figure}[ht]
   \epsfxsize=5.0in 
\centerline{\epsffile{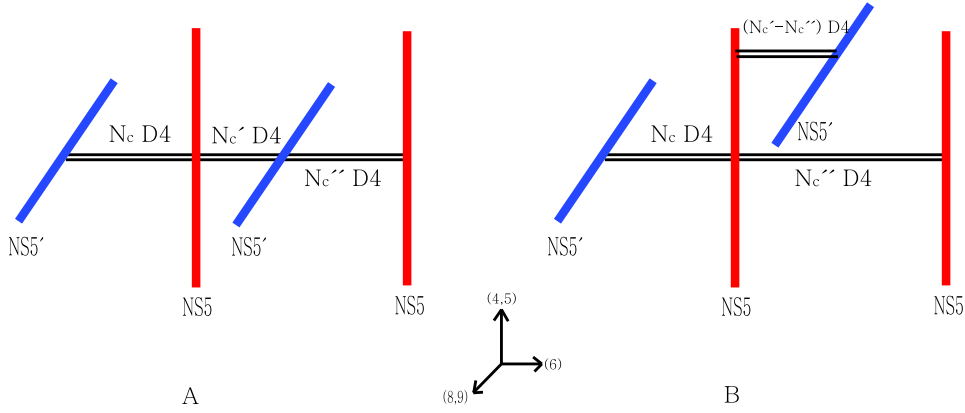}}
   \caption[FIG. \arabic{figure}.]{ 
The  ${\cal N}=1$ supersymmetric 
electric brane configuration for the gauge group $SU(N_c) \times
SU(N_c') \times SU(N_c'')$ and  bifundamentals $F, \widetilde{F}, G$
and $\widetilde{G}$  with vanishing(1A) and 
nonvanishing(1B) mass
for the bifundamentals $F$ and $\widetilde{F}$. 
The $N_c'$ D4-branes in 1A are decomposed into 
$(N_c'-N_c'')$ D4-branes which are moving to $+v$ direction in 1B 
and $N_c''$ D4-branes which are recombined with those D4-branes
connecting between $NS5_R'$-brane and $NS5_R$-brane in 1B. }
\end{figure}

The gauge couplings of $SU(N_c)$, $SU(N_c')$ and $ SU(N_c'')$
are given by a string coupling constant $g_s$, a string scale $\ell_s$ 
and the $x^6$ coordinates $y_i$ for three NS-branes through
\bea
g_1^2 =\frac{g_s \ell_s}{y_1}, \qquad 
g_2^2 = \frac{g_s \ell_s}{y_2}, \qquad
g_3^2=\frac{g_s \ell_s}{y_3}.
\nonu
\eea
For example, as $y_3$ goes to $\infty$, the  
$SU(N_c'')$ gauge group becomes a
global symmetry and the theory leads to SQCD with the gauge group
$SU(N_c) \times SU(N_c')$ and $N_c''$ flavors 
in the fundamental representation.

There is no superpotential in Figure 1A since the $NS5_L$-brane is 
perpendicular to two NS5'-branes and the $NS5_R'$-brane is
perpendicular to two NS5-branes. Let us deform this theory.
Displacing the two NS5'-branes relative each other in the 
\bea
v \equiv
x^4 + i x^5
\nonu
\eea 
direction corresponds to turning on a quadratic
mass-deformed superpotential
for the fields $F$ and $\widetilde{F}$ as follows:
\bea
W = m F \widetilde{F} \equiv m \Phi'
\label{mass}
\eea
where 
the first gauge group indices in $F$ and $\widetilde{F}$ 
are contracted, each second gauge group index in them is encoded in 
$\Phi'$ and the mass $m$ is given by
\bea
m =\frac{\Delta x}{2\pi \alpha'} = \frac{\Delta x}{\ell_s^2}.
\label{m}
\eea

The gauge-singlet $\Phi'$ for the first dual gauge group is in the 
adjoint representation for the second dual gauge group, 
i.e., ${(\bf 1, (N_c'-N_c'')^2-1,1)  \oplus (1,1,1)}$ 
under the dual gauge group (\ref{dual}). 
Then the $\Phi'$ is a $(N_c'-N_c'') \times (N_c'-N_c'')$ matrix.
The $NS5_R'$-brane together with $(N_c'-N_c'')$-color D4-branes 
is moving to the $+v$ direction  for
fixed other branes during this mass deformation. 
In other words, the $N_c''$ D4-branes among $N_c'$ D4-branes 
are not participating in 
the mass deformation.
Then the $x^5$ coordinate($\equiv x$) 
of $NS5_L'$-brane is equal to
zero
while the $x^5$ coordinate of $NS5_R'$-brane is given by 
$\Delta x$.
Giving an expectation value to the meson field $\Phi'$
corresponds to recombination of $N_c$- and $N_c'$- color 
D4-branes, which will become $N_c$-color D4-branes
in Figure 1A such that they are suspended between 
the $NS5_L'$-brane and the $NS5_R'$-brane 
and pushing them into the 
\bea
w \equiv x^8 + i
x^9
\nonu
\eea
direction. We assume that the number of colors satisfies
\bea
N_c' \geq N_c \geq N_c''.
\nonu
\eea

Now 
we draw this brane configuration in Figure 1B for nonvanishing mass
for the fields $F$ and $\widetilde{F}$. 

\subsection{Magnetic theory}

By applying the Seiberg dual to the $SU(N_c)$ factor in 
(\ref{gaugegroup}), the two $NS5_{L,R}'$-branes can be located at the
inside of the two NS5-branes, as in Figure 2.
Starting from Figure 1A and moving the $NS5_R'$-brane with $(N_c'-N_c'')$
D4-branes 
to the $+v$ direction leading to Figure 1B 
and interchanging the $NS5_L'$-brane and the $NS5_L$-brane,
one obtains the Figure 2A.
Before arriving at the Figure 2A, there exists an intermediate 
step where the $(N_c'-N_c)$ D4-branes are connecting between the 
$NS5_L$-brane and the  $NS5_L'$-brane,  
$(N_c'-N_c'')$ D4-branes connecting between the  $NS5_L'$-brane and   
$NS5_R'$-brane, and $N_c''$ D4-branes between the $NS5_L'$-brane and
the $NS5_R$-brane. By introducing $-N_c''$ D4-branes and $-N_c''$ 
anti-D4-branes  between the  $NS5_L$-brane and   
$NS5_L'$-brane, reconnecting the former with  
the $N_c'$ D4-branes connecting between  
$NS5_L$-brane 
and the $NS5_L'$-brane (therefore $N_c'-N_c''$ D4-branes)
and moving those combined
$(N_c'-N_c'')$ 
D4-branes
to $+v$-direction, 
one gets the final Figure 2A where we are left with 
$(N_c-N_c'')$ 
anti-D4-branes between the $NS5_L$-brane and   
$NS5_L'$-brane.

When two NS5'-branes in Figure 2A are close to each other, then 
it leads to Figure 2B by realizing that the number of $(N_c'-N_c'')$
D4-branes connecting between $NS5_L$-brane and $NS5_R'$-brane can
be rewritten as $(N_c-N_c'')$ plus $\widetilde{N}_c$.
If we ignore $N_c''$ D4-branes and $NS5_R$-brane  from Figure 2B, then
the brane configuration becomes the one in \cite{GK}.

\begin{figure}[ht]
   \epsfxsize=5.0in 
\centerline{\epsffile{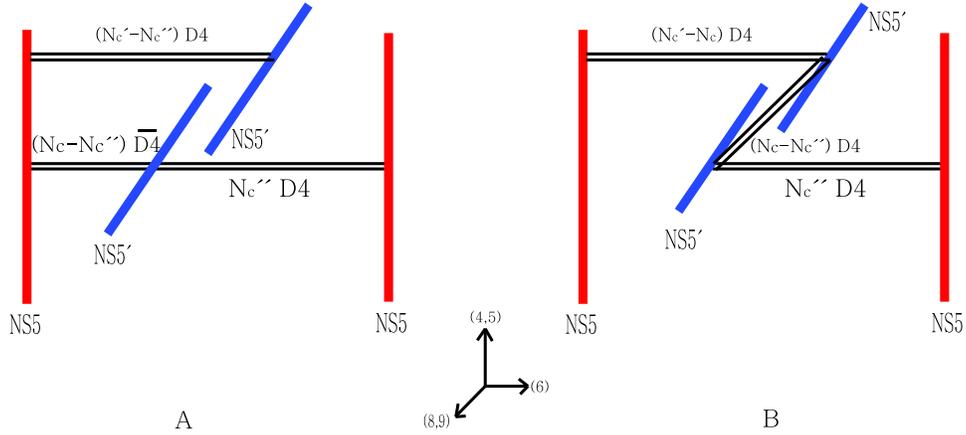}}
   \caption[FIG. \arabic{figure}.]{ 
The  ${\cal N}=1$ magnetic brane configuration for the gauge group 
$SU(\widetilde{N}_c=N_c'-N_c) 
\times SU(N_c') \times SU(N_c'')$ 
corresponding to Figure 1B with D4-
and $\overline{D4}$-branes(2A) and with 
a misalignment between D4-branes(2B) when the NS5'-branes are close to
each other. The number of tilted D4-branes in 2B can be written as
$N_c-N_c''=(N_c'-N_c'')-\widetilde{N}_c$. }
\end{figure}

The dual gauge group is given by 
\bea
SU(\widetilde{N}_c=N_c'-N_c) \times SU(N_c') \times SU(N_c'').
\label{dual}
\eea
The matter contents are the field $f$ 
 charged under
$({\bf \widetilde{N}_c, \overline{N_c'}, 1})$, a field $g$ charged under
$({\bf 1, N_c', \overline{N_c''}})$, and their conjugates 
$\widetilde{f}$ and $\widetilde{g}$ under the dual gauge group
(\ref{dual})
and  
the gauge-singlet $\Phi'$ for the first dual gauge group in the 
adjoint representation for the second dual gauge group, 
i.e., ${(\bf 1, (N_c'-N_c'')^2-1,1)  \oplus (1,1,1)}$ under the 
dual gauge group (\ref{dual}).

The cubic superpotential with the mass term (\ref{mass}) in the dual
theory is given by
\bea
W_{dual} = \Phi' f \widetilde{f} + m \Phi'. 
\label{Wdual}
\eea
Here the magnetic fields $f$ and $\widetilde{f}$  
correspond to 4-4 strings connecting 
the $\widetilde{N}_c$-color D4-branes(that are 
connecting between the $NS5_L$-brane
and the $NS5_R'$-brane in Figure 2B) with $N_c'$-flavor 
D4-branes(that are 
a combination of three different D4-branes in Figure 2B).
Among these $N_c'$-flavor D4-branes, only the strings ending on
the upper $(N_c'-N_c)$ D4-branes and 
on the tilted middle $(N_c-N_c'')$ 
D4-branes in Figure 2B enter the cubic superpotential term. 
Although the $(N_c'-N_c'')$ D4-branes in Figure 2A cannot move any
directions,
the tilted $(N_c-N_c'')$-flavor D4-branes 
can move $w$ direction in Figure 2B.
The remaining upper $\widetilde{N}_c$ D4-branes are fixed also and cannot 
move any direction. Note that 
there is a decomposition 
\bea
(N_c'-N_c'')=(N_c-N_c'')+\widetilde{N}_c.
\nonu
\eea 

The brane configuration for zero mass for the bifundamental,
which has only a cubic superpotential,
can be obtained from Figure 2A by moving
the upper $NS5_R'$-brane together with $(N_c'-N_c'')$ color D4-branes 
into the origin $v=0$.
Then the number of dual colors for D4-branes 
becomes  
$\widetilde{N}_c$ between $NS5_L$-brane and $NS5_L'$-brane 
and
$N_c'$ between two NS5'-branes
as well
as $N_c''$ D4-branes between $NS5_R'$-brane and $NS5_R$-brane.
Or starting from Figure 1A and moving the $NS5_L$-brane to the left all the
way past the $NS5_L'$-brane,
one also obtains the corresponding magnetic brane configuration
for massless case.

The brane configuration in Figure 2A is stable as long as the
distance $\Delta x$ between the upper NS5'-brane and 
the lower NS5'-brane is large, as
in \cite{GK}. If they are close to each other, then this brane
configuration is unstable to decay and leads to 
the brane configuration in Figure
2B.
One can regard these brane configurations as particular states in the
magnetic gauge theory with the gauge group (\ref{dual}) and
superpotential (\ref{Wdual}).
The   $(N_c'-N_c''-\widetilde{N}_c)$ flavor D4-branes of 
straight brane configuration
of
Figure 2B  bend due to the fact that there exists an attractive
gravitational interaction
between those flavor D4-branes and $NS5_L$-brane from the DBI action, by
following the procedure of \cite{GK}, as long as the distance $y_3$
goes to $\infty$ because the presence of an extra $NS5_R$-brane does
not affect the DBI action. 
For the finite and small $y_3$, the careful analysis for DBI action is
needed in
order to obtain the bending curve connecting  two NS5'-branes.  

When the upper NS5'-brane(or $NS5_R'$-brane) 
is replaced by coincident $(N_c'-N_c'')$ 
D6-branes and 
the $NS5_R$ is rotated by an angle $\frac{\pi}{2}$ in the $(v,w)$
plane in Figure 2B, this brane configuration reduces to the one 
found in \cite{Ahn07-3} where the gauge group was given by 
$SU(n_f+n_c'-n_c) \times SU(n_c')$ 
with $n_f$ multiplets,  $n_f'$ multiplets, flavor singlets and gauge 
singlets. 
Then the present number $(N_c'-N_c'')$ corresponds to the $n_f$, the
number $N_c$ corresponds to $n_c$ and 
the number $N_c''$ corresponds to the $n_c'$ of \cite{Ahn07-3}.
Note that the number of D4-branes touching $NS5_R'$-brane in Figure 2B
is equal to $(N_c'-N_c'')$.

The quantum corrections can be understood for small $\Delta x$ by 
using the low energy field theory on the branes.
The low energy dynamics of the magnetic brane configuration 
can be described by the ${\cal N}=1$ supersymmetric gauge theory
with gauge group (\ref{dual})
and the gauge couplings for the three gauge group factors are
given by
\bea
g_{1,mag}^2  = \frac{g_s \ell_s}{y_1}, \qquad 
g_{2,mag}^2 = \frac{g_s \ell_s}{(y_2-y_1)}, \qquad
g_{3,mag}^2  = \frac{g_s \ell_s}{y_3}.
\nonu
\eea
The dual gauge theory has  an adjoint $\Phi'$ of $SU(N_c')$ and 
bifundamentals $f, \widetilde{f}, g$ and $\widetilde{g}$ under the dual gauge
group (\ref{dual}) and the superpotential 
corresponding to Figures 2A and 2B is given by 
\bea
W_{dual} = h \Phi' f \widetilde{f} - h \mu^2 \Phi', \qquad h^2 = g_{2,
  mag}^2,
\qquad \mu^2 = -\frac{\Delta x}{ 2\pi g_s \ell_s^3}.
\nonu
\eea
Then $ f \widetilde{f}$ is a $\widetilde{N}_c \times \widetilde{N}_c$ 
matrix where the second gauge group indices for $f$ and $\widetilde{f}$ 
are contracted with those
of $\Phi'$ while $\mu^2$ is a 
$(N_c'-N_c'') \times (N_c'-N_c'')$ matrix.
Although the field $f$ itself is an antifundamental in the second gauge
group
which is a different  
representation for the usual standard quark
coming from D6-branes,
the product $f \widetilde{f}$ has the same representation for the 
product of quarks
and moreover, 
the second gauge group indices for the field $\Phi'$ play the
role of the flavor indices, as in comparison with the brane
configuration in the presence of D6-branes before.

Therefore, the F-term equation, the derivative $W_{dual}$ with respect to the
meson field $\Phi'$ cannot be satisfied if the $(N_c'-N_c'')$ exceeds
$\widetilde{N}_c$.
So the supersymmetry is broken.   
That is, 
there exist three equations from F-term conditions:
\bea
f\widetilde{f} -\mu^2 =0, \qquad  \mbox{and} 
\qquad \Phi' f =0=\widetilde{f} \Phi'.
\nonu
\eea
Then the solutions for these
are given by 
\bea
<f>   = 
\left(
\begin{array}{c}
\mu  {\bf 1}_{\widetilde{N}_c}  \\
0
\end{array}
\right), 
\qquad
<\widetilde{f}>   = 
\left(
\begin{array}{cc}
\mu  {\bf 1}_{\widetilde{N}_c} & 0  \\
\end{array}
\right), 
\qquad
<\Phi'> =
 \left(
\begin{array}{cc}
0  & 0  \\
0 & \Phi'_0  {\bf 1}_{(N_c'-N_c''-\widetilde{N}_c)} 
\end{array}
\right) 
\label{point}
\eea
where the zero of $<f>$ is a $
(N_c'-N_c''-\widetilde{N}_c) \times \widetilde{N}_c$ 
matrix, the zero of $<\widetilde{f}>$ is a
$\widetilde{N}_c \times (N_c'-N_c''-\widetilde{N}_c) $ matrix and 
the zeros of $<\Phi'>$ are $\widetilde{N}_c \times \widetilde{N}_c$,
$\widetilde{N}_c \times (N_c'-N_c''-\widetilde{N}_c)$, 
and $(N_c'-N_c''-\widetilde{N}_c) \times
\widetilde{N}_c$ matrices.
Then one can expand these fields around on a point (\ref{point}), as
in \cite{ISS,Shih,Ahn07-2,Ahn07-1,Ahn07} 
and one arrives at the relevant superpotential
up to quadratic order in the fluctuation. 
At one loop, the effective potential $V_{eff}^{(1)}$ for $\Phi'_0$
leads to the positive value for $m_{\Phi'_0}^2$ implying that these
vacua are stable.

\subsection{Other magnetic theory-I}

Let us consider other magnetic theory for the same undeformed electric theory
given in the subsection 2.1. Here we consider the different mass deformation.
By applying the Seiberg dual to the $SU(N_c')$ factor in 
(\ref{gaugegroup}), the two $NS5_{L,R}'$-branes can be located at the
left hand side of the two NS5-branes, as in Figure 4.

\begin{figure}[ht]
   \epsfxsize=5.0in 
\centerline{\epsffile{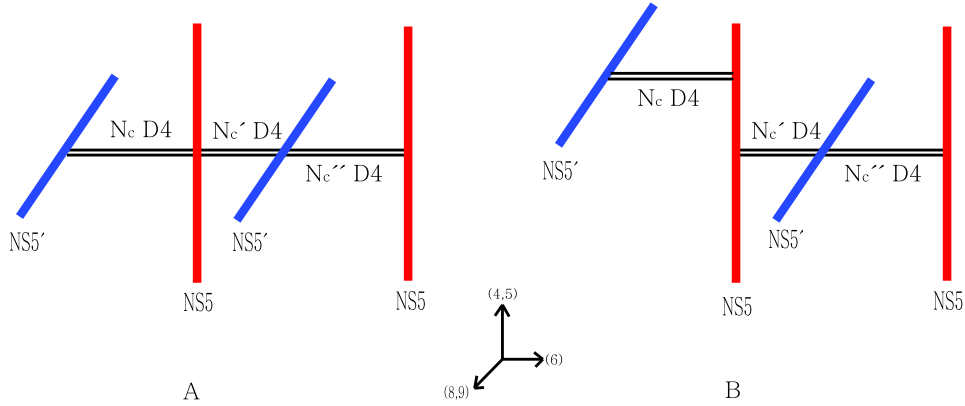}}
   \caption[FIG. \arabic{figure}.]{ 
The  ${\cal N}=1$ supersymmetric 
electric brane configuration for the gauge group $SU(N_c) \times
SU(N_c') \times SU(N_c'')$ 
and  bifundamentals $F, \widetilde{F}, G$ and $\widetilde{G}$  
with vanishing(3A) which is identical to Figure 1A and 
nonvanishing(3B) mass
for the bifundamentals $F$ and $\widetilde{F}$.
This deformation is different from the one (\ref{mass}) given previously. 
In 3B, the $NS5_L'$-brane together with $N_c$ D4-branes 
is moving to $+v$ direction.}
\end{figure}

By starting from Figure 3A which is the same as Figure 1A
and moving the $NS5_L'$-brane with $N_c$
D4-branes 
to the $+v$ direction leading to 
Figure 3B and interchanging the $NS5_L$-brane and the $NS5_R'$-brane,
one obtains the Figure 4A.
Before arriving at the Figure 4A, there exists an intermediate 
step where the $N_c$ D4-branes are connecting between the 
$NS5_L'$-brane and the $NS5_R'$-brane,  
$(N_c''-N_c'+N_c)$ D4-branes are connecting between the $NS5_R'$-brane and   
$NS5_L$-brane, and $N_c''$ D4-branes are suspended 
between the $NS5_L$-brane and
the $NS5_R$-brane. 
By moving the combined
$N_c$ D4-branes, obtained from reconnection of those D4-branes between 
$NS5_L'$-brane and the $NS5_R'$-brane and those D4-branes
 between the $NS5_R'$-brane and   
$NS5_L$-brane(therefore between the $NS5_L'$-brane and the $NS5_L$-brane), 
to $+v$-direction, 
one gets the final Figure 4A where we are left with 
$(N_c'-N_c'')$ 
anti-D4-branes between the $NS5_R'$-brane and   
$NS5_L$-brane.
We assume  that the number of colors satisfies
\bea
N_c+N_c'' \geq N_c' \geq N_c''.
\nonu
\eea
When two NS5'-branes in Figure 4A are close to each other, then 
it leads to Figure 4B
 by realizing that the number of $N_c$
D4-branes connecting between $NS5'_L$-brane and $NS5_L$-brane can
be rewritten as $(N_c'-N_c'')$ plus $\widetilde{N}_c'$.

\begin{figure}[ht]
   \epsfxsize=5.0in 
\centerline{\epsffile{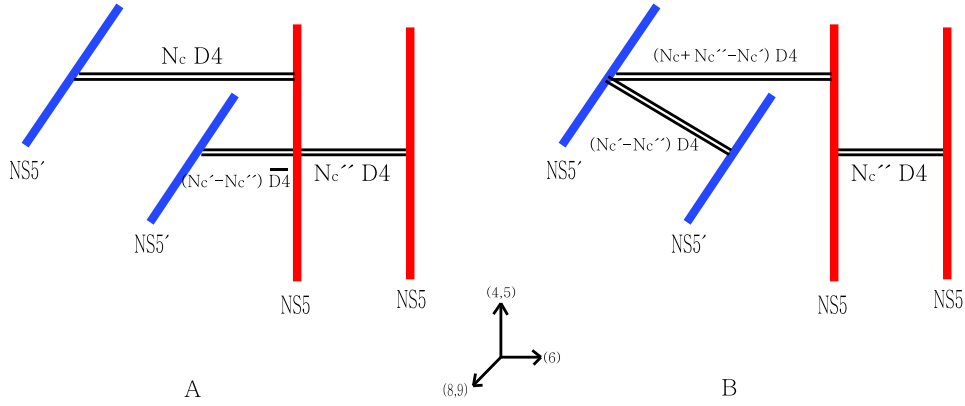}}
   \caption[FIG. \arabic{figure}.]{ 
The  ${\cal N}=1$ 
magnetic brane configuration for the gauge group $SU(N_c) \times 
SU(\widetilde{N}_c'=N_c+N_c''-N_c')
\times SU(N_c'')$ 
corresponding to Figure 3B with D4-
and $\overline{D4}$-branes(4A) and with 
a misalignment between D4-branes(4B) when the NS5'-branes are close to
each other. The number of tilted D4-branes is equal to $N_c'-N_c''=
N_c-\widetilde{N}_c'$ in 4B.  }
\end{figure}

The dual gauge group is
\bea
SU(N_c) \times SU(\widetilde{N}_c'=N_c+N_c''-N_c') \times SU(N_c'').
\label{dualg}
\eea
The matter contents are the field $f$ 
 charged under
$({\bf N_c, \overline{\widetilde{N}_c'}, 1})$, a field $g$ charged under
$({\bf 1, \widetilde{N}_c', \overline{N_c''}})$, and their conjugates 
$\widetilde{f}$ and $\widetilde{g}$ under the dual gauge group
(\ref{dualg})
and  
the gauge-singlet $\Phi$ for the second dual gauge group in the 
adjoint representation for the first dual gauge group, 
i.e., ${(\bf N_c^2-1,1,1)  \oplus (1,1,1)}$ under the 
dual gauge group (\ref{dualg}).
Then the $\Phi$ is a $N_c \times N_c$ matrix.
All the $N_c$ D4-branes participate in the mass deformation.

The cubic superpotential with the mass term  in the dual
theory \footnote{One can also construct the mass deformation by
  rotating $NS5_R$-brane and  moving
it to $+v$ direction. Then the brane configuration will look like as the
Figure 15. } is given by
\bea
W_{dual} = \Phi f \widetilde{f} + m \Phi 
\label{dualW}
\eea
where we define $\Phi$ as $\Phi \equiv F \widetilde{F}$, 
the second gauge group indices in $F$ and $\widetilde{F}$ 
are contracted, each first gauge group index in them is encoded in 
$\Phi$ and the mass $m$ is given by (\ref{m}) where 
$\Delta x$ refers to the distance between two NS5'-branes along the
$x^5$ direction in Figure 4A.
Let us emphasize that although the $\Phi$ which has first gauge group
indices 
looks similar to the
previous $\Phi'$ which has second gauge group indices 
in (\ref{mass}), the group indices are different.
Here the magnetic fields $f$ and $\widetilde{f}$  
correspond to 4-4 strings connecting 
the $\widetilde{N}_c'$-color D4-branes(that are 
connecting between the $NS5_L'$-brane
and the $NS5_L$-brane in Figure 4B) with $N_c$-flavor 
D4-branes(which  are realized 
as corresponding D4-branes in Figure 4A).
Although the $N_c$ D4-branes in Figure 4A cannot move any
directions,
the tilted $(N_c'-N_c'')$-flavor D4-branes can move 
$w$ direction in Figure 4B.
The remaining upper $\widetilde{N}_c'$ D4-branes are fixed also and cannot 
move any direction. Note that 
there is a decomposition 
\bea
N_c=(N_c'-N_c'')+\widetilde{N}_c'.
\nonu
\eea 

The brane configuration for zero mass for the bifundamental,
which has only a cubic superpotential,
can be obtained from Figure 4A by moving
the upper NS5'-brane(or $NS5_L'$-brane) together with $N_c$ color D4-branes 
into the origin $v=0$.
Then the number of dual colors for D4-branes 
becomes $N_c$ between two NS5'-branes, 
 $\widetilde{N}_c'$ between $NS5_R'$-brane and $NS5_L$-brane
and $N_c''$ 
between $NS5_L$-brane and $NS5_R$-brane.
Or starting from Figure 3A and moving the $NS5_L$-brane to the right all the
way past the $NS5_R'$-brane,
one also obtains the corresponding magnetic brane configuration
for massless case.

The brane configuration in Figure 4A is stable as long as the
distance $\Delta x$ between the upper NS5'-brane and 
the lower NS5'-brane is large. If they are close to each other, then this brane
configuration is unstable to decay to 
the brane configuration in Figure
4B.
One can regard these brane configurations as particular states in the
magnetic gauge theory with the gauge group (\ref{dualg}) and
superpotential (\ref{dualW}).
The $(N_c-\widetilde{N}_c')$ flavor D4-branes of 
straight brane configuration
of
Figure 4B  bend since there exists an attractive
gravitational interaction
between those flavor D4-branes and $NS5_L$-brane from the DBI action, 
as long as the distance $y_3$
is large because the presence of an extra $NS5_R$-brane does
not affect the DBI action. 
For the finite and small $y_3$, the careful analysis for DBI action is
needed in
order to obtain the bending curve connecting  two NS5'-branes.  
Or if $y_3$ goes to zero, then this extra $NS5_R$-brane plays the role
of enhancing the strength for the  
NS5-branes and will affect both the energy of bending curve, 
$E_{curved}$, 
which is proportional to $\frac{1}{l}$ with $l \equiv \sqrt{k} \ell_s$ where
$k$ is the number of NS5-branes and $\Delta x$ which depends on
both $\frac{1}{l}$ and $l$ \cite{GK}.

When the upper NS5'-brane(or $NS5_L'$-brane) 
is replaced by coincident $N_c$(that is equal to the number of
D4-branes touching the $NS5_L'$-brane)
D6-branes, this brane configuration looks similar to the one 
found in \cite{Ahn07-3} where the gauge group was given by 
$SU(n_f'+n_c-n_c') \times SU(n_c)$ 
with $n_f$ multiplets, $n_f'$ multiplets, flavor singlets and gauge singlets. 
Then the present $N_c$ corresponds to the $n_f'$, the number $N_c'$
corresponds to $n_c'$, and 
$N_c''$ corresponds to the $n_c$ of \cite{Ahn07-3}. 

The low energy dynamics of the magnetic brane configuration 
can be described by the ${\cal N}=1$ supersymmetric gauge theory
with gauge group (\ref{dualg})
and the gauge couplings for the three gauge group factors are
given by
\bea
g_{1,mag}^2  = \frac{g_s \ell_s}{(y_1-y_2)}, \qquad 
g_{2,mag}^2 = \frac{g_s \ell_s}{y_2}, \qquad
g_{3,mag}^2  = \frac{g_s \ell_s}{(y_2+ y_3)}.
\nonu
\eea

The dual gauge theory has  an adjoint $\Phi$ of $SU(N_c)$ and 
bifundamentals $f, \widetilde{f}, g$ and $\widetilde{g}$ under the dual gauge
group (\ref{dualg}) and the superpotential 
corresponding to Figures 4A and 4B is given by 
\bea
W_{dual} = h \Phi f \widetilde{f} - h \mu^2 \Phi, \qquad h^2 = g_{1,
  mag}^2,
\qquad \mu^2 = -\frac{\Delta x}{ 2\pi g_s \ell_s^3}.
\nonu
\eea
Then $ f \widetilde{f}$ is a $\widetilde{N}_c' \times \widetilde{N}_c'$ 
matrix where the first gauge group indices for $f$ and $\widetilde{f}$ 
are contracted with those
of $\Phi$ while $\mu^2$ is a 
$N_c \times N_c$ matrix.
The product $f \widetilde{f}$ has the same representation for the 
product of quarks
and moreover, 
the first gauge group indices for the field $\Phi$ play the
role of the flavor indices when there are D6-branes before.

Therefore, the F-term equation, the derivative $W_{dual}$ with respect to the
meson field $\Phi$ cannot be satisfied if the $N_c$ exceeds
$\widetilde{N}_c'$.
So the supersymmetry is broken.   
That is, 
there exist three equations from F-term conditions:
$
f\widetilde{f} -\mu^2 =0$ and $ \Phi f =0=\widetilde{f} \Phi$.
Then the solutions for these
are given by 
\bea
<f>   = 
\left(
\begin{array}{c}
\mu  {\bf 1}_{\widetilde{N}_c'}  \\
0
\end{array}
\right), 
\qquad
<\widetilde{f}>   = 
\left(
\begin{array}{cc}
\mu  {\bf 1}_{\widetilde{N}_c'} & 0  \\
\end{array}
\right), 
\qquad
<\Phi> =
 \left(
\begin{array}{cc}
0  & 0  \\
0 & \Phi_0  {\bf 1}_{(N_c-\widetilde{N}_c')} 
\end{array}
\right) 
\label{point1}
\eea
where the zero of $<f>$ is a $
(N_c-\widetilde{N}_c') \times \widetilde{N}_c'$ 
matrix, the zero of $<\widetilde{f}>$ is a
$\widetilde{N}_c' \times (N_c-\widetilde{N}_c') $ matrix and 
the zeros of $<\Phi>$ are $\widetilde{N}_c' \times \widetilde{N}_c'$,
$\widetilde{N}_c' \times 
(N_c-\widetilde{N}_c')$ and $(N_c-\widetilde{N}_c') \times
\widetilde{N}_c'$ 
matrices.
Then one can expand these fields around on a point (\ref{point1}), as
in \cite{ISS,Shih} and one arrives at the relevant superpotential
up to quadratic order in the fluctuation. 
At one loop, the effective potential $V_{eff}^{(1)}$ for $\Phi_0$
leads to the positive value for $m_{\Phi_0}^2$ implying that these
vacua are stable.

\subsection{Other magnetic theory-II}

One can think of the following dual gauge group
\bea
SU(N_c) \times SU(N_c') \times SU(\widetilde{N}_c''=N_c'-N_c'')
\label{newdual}
\eea
by performing the magnetic dual for the last gauge group 
in (\ref{gaugegroup}).
The electric brane configuration can be given in terms of Figure 1A or
Figure 1A with an exchange between NS5-brane and NS5'-brane. Then for
the latter, the
resulting brane configuration is given by $NS5_L$-brane,
$NS5_L'$-brane, $NS5_R$-brane, and $NS5'_R$-brane from the left to the
right in the $x^6$ direction. 

In order to obtain the above dual gauge
group, we need to interchange between the $NS5_R$-brane and the 
$NS5_R'$-brane, as we did before. One can do this either by following
the previous procedure or by looking at the Figure 2 from the negative
$w$ direction which is an opposite viewpoint, compared with Figure 2.
In other words, we are looking at the Figure 2 from the other side of
$w$.
Then the resulting brane configuration in this case can be obtained by 
taking a reflection for all the NS-branes, D4-branes and anti
D4-branes with respect to the $NS5_L$-brane(rotating them to the left
for fixed $NS5_L$-brane) in Figure 2A and Figure 2B. 
Then  the ${\cal N}=1$ magnetic brane configuration for the gauge group 
$SU(N_c) 
\times SU(N_c') \times SU(\widetilde{N}_c''=N_c'-N_c'')$ 
corresponds to  the Figure 5A' with D4-
and $\overline{D4}$-branes and the Figure 5B' with 
a misalignment between D4-branes when the NS5'-branes are close to
each other. The number of tilted D4-branes in 5B' can be written as
$N_c''-N_c=(N_c'-N_c)-\widetilde{N}_c''$. We do not present 
the Figures 5A' and 5B' here.

Let us consider other magnetic theory for the same electric theory
given in the subsection 2.1 with Figure 1A.
By applying the Seiberg dual to the $SU(N_c'')$ factor in 
(\ref{gaugegroup}) and 
interchanging the $NS5_R'$-brane and the $NS5_R$-brane,
one obtains the Figure 5A''.
Before arriving at the Figure 5A'', there exists an intermediate 
step where 
$N_c$ D4-branes between
$NS5_L'$-brane and the $NS5_R$-brane,
the $N_c'$ D4-branes are connecting between the 
$NS5_L$-brane and the  $NS5_R$-brane, and  
$(N_c'-N_c'')$ D4-branes are connecting between the  $NS5_R$-brane and   
$NS5_R'$-brane. 
By rotating $NS5_L$-brane by an angle $\frac{\pi}{2}$ which will
become $NS5_M'$-brane, 
moving it with the $(N_c'-N_c)$ D4-branes 
to $+v$ direction where we introduce $(N_c'-N_c)$ D4-branes and
$(N_c'-N_c)$ anti D4-branes between the $NS5_R$-brane and the 
$NS5_R'$-brane, 
one gets the final Figure 5A'' where we are left with 
$(N_c''-N_c)$ 
anti-D4-branes between the NS5-brane and   
the $NS5_R'$-brane.
When two NS5'-branes in Figure 5A'' are close to each other, then 
it leads to Figure 5B''
 by realizing that the number of $(N_c'-N_c)$
D4-branes connecting between $NS5_M'$-brane and NS5-brane can
be rewritten as $(N_c''-N_c)$ plus $\widetilde{N}_c''$.

\begin{figure}[ht]
   \epsfxsize=5.0in 
\centerline{\epsffile{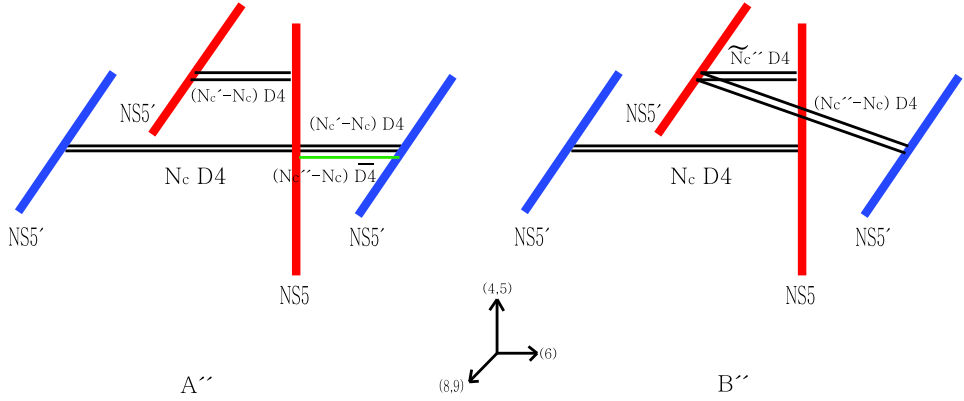}}
   \caption[FIG. \arabic{figure}.]{ 
The  ${\cal N}=1$ magnetic brane configuration for the gauge group 
$SU(N_c) 
\times SU(N_c') \times SU(\widetilde{N}_c''=N_c'-N_c'')$ 
with D4-
and $\overline{D4}$-branes(5A'') and with 
a misalignment between D4-branes(5B'') when the NS5'-branes are close to
each other. The number of tilted D4-branes in 5B'' can be written as
$N_c''-N_c=(N_c'-N_c)-\widetilde{N}_c''$. The deformation is related
to
the bifundamentals $G$ and $\widetilde{G}$.  }
\end{figure}

The brane configuration in Figure 5A'' is stable as long as the
distance $\Delta x$ between the upper NS5'-brane and 
the lower NS5'-brane(or $NS5_R'$-brane) 
is large. If they are close to each other, then this brane
configuration is unstable to decay to 
the brane configuration in Figure
5B''.
One can regard these brane configurations as particular states in the
magnetic gauge theory with the gauge group and
superpotential.
The   $(N_c'-N_c-\widetilde{N}_c'')$ flavor D4-branes of 
straight brane configuration
of
Figure 5B''  bend since there exists an attractive
gravitational interaction
between those flavor D4-branes and NS5-brane from the DBI action. 
As mentioned in \cite{Ahn07-5},
the two NS5'-branes are located at different side of NS5-brane in
Figure 5B'' and the DBI action computation for this bending curve
should be taken into account. 

The matter contents are the field $f$ 
 charged under
$( {\bf N_c, \overline{N_c'}, 1})$, a field $g$ charged under
$({\bf 1, N_c', \overline{\widetilde{N}_c''}})$ 
 and their conjugates 
$\widetilde{f}$ and $\widetilde{g}$
under the dual gauge group
(\ref{newdual})
and  
the gauge-singlet $\Phi'$ which is in the 
adjoint representation for the second dual gauge group, 
in other words,   
$ ({ \bf   1,  (N_c'-N_c)^2-1,1})  \oplus  ({\bf 1,1,1})$ under the 
dual gauge group (\ref{newdual}).
Then the $\Phi'$ is a $(N_c'-N_c) \times (N_c'-N_c)$ matrix.
Only $(N_c'-N_c)$ D4-branes can participate in the mass deformation.

The cubic superpotential with the mass term
is given by
\bea
W_{dual} = \Phi' g \widetilde{g} + m \Phi'
\label{ssuper}
\eea
where we define $\Phi'$ as $\Phi' \equiv G \widetilde{G}$ and 
the third gauge group indices in $G$ and $\widetilde{G}$ 
are contracted, each second gauge group index in them is encoded in 
$\Phi'$. 
Here the magnetic fields $g$ and $\widetilde{g}$  
correspond to 4-4 strings connecting 
the $\widetilde{N}_c''$-color D4-branes(that are 
connecting between the $NS5_M'$-brane
and the NS5-brane in Figure 5B'') with $N_c'$-flavor 
D4-branes.
Among these $N_c'$-flavor D4-branes, only the strings ending on
the upper $(N_c'-N_c'')$ D4-branes and 
on the tilted middle $(N_c''-N_c)$ 
D4-branes in Figure 5B'' enter the cubic superpotential term. 
Although the $(N_c'-N_c)$ D4-branes in Figure 5A'' cannot move any
directions,
the tilted $(N_c''-N_c)$-flavor D4-branes can move $w$ direction.
The remaining upper $\widetilde{N}_c''$ D4-branes are fixed also and cannot 
move any direction. 
Note that 
there is a decomposition 
\bea
(N_c'-N_c)=(N_c''-N_c)+\widetilde{N}_c''.
\nonu
\eea 

The brane configuration for zero mass for the bifundamental,
which has only a cubic superpotential,
can be obtained from Figure 5A'' by moving
the upper  NS5'-brane together with $(N_c'-N_c)$ color D4-branes 
into the origin $v=0$.
Then the number of dual colors for D4-branes 
becomes $N_c$ between the $NS5_L'$-brane and the $NS5_M'$-brane, 
$N_c'$ between the $NS5_M'$-brane and the  NS5-brane
and $\widetilde{N}_c''$ 
between NS5-brane and $NS5_R'$-brane.
Or starting from Figure 1A and moving the 
$NS5_R'$-brane to the right all the
way past the $NS5_R$-brane,
one also obtains the corresponding magnetic brane configuration
for massless case.

The low energy dynamics of the magnetic brane configuration 
can be described by the ${\cal N}=1$ supersymmetric gauge theory
with gauge group (\ref{newdual})
and the gauge couplings for the three gauge group factors are
given by
\bea
g_{1,mag}^2  = \frac{g_s \ell_s}{y_1}, \qquad 
g_{2,mag}^2 = \frac{g_s \ell_s}{(y_2-y_3)}, \qquad
g_{3,mag}^2  = \frac{g_s \ell_s}{y_3}.
\nonu
\eea
The dual gauge theory has  an adjoint $\Phi'$ of $SU(N_c')$ and 
bifundamentals $f, \widetilde{f}, g$ and $\widetilde{g}$ under the dual gauge
group (\ref{newdual}) and the superpotential 
corresponding to Figures 5A'' and 5B'' is given by 
\bea
W_{dual} = h \Phi' g \widetilde{g} - h \mu^2 \Phi', \qquad h^2 = g_{2,
  mag}^2,
\qquad \mu^2 = -\frac{\Delta x}{ 2\pi g_s \ell_s^3}.
\nonu
\eea
Then $ g \widetilde{g}$ is a $\widetilde{N}_c'' \times \widetilde{N}_c''$ 
matrix where the second gauge group indices for $g$ and $\widetilde{g}$ 
are contracted with those
of $\Phi'$ while $\mu^2$ is a 
$(N_c'-N_c) \times (N_c'-N_c)$ matrix.
The product $g \widetilde{g}$ has the same representation for the 
product of quarks
and moreover, 
the second gauge group indices for the field $\Phi'$ play the
role of the flavor indices.

Therefore, the F-term equation, the derivative $W_{dual}$ with respect to the
meson field $\Phi'$ cannot be satisfied if the $(N_c'-N_c)$ exceeds
$\widetilde{N}_c''$.
So the supersymmetry is broken.   
That is, 
there exist three equations from F-term conditions:
$
g\widetilde{g} -\mu^2 =0$ and $ \Phi' g =0=\widetilde{g} \Phi'$.
Then the solutions for these
are given by 
\bea
<g>   = 
\left(
\begin{array}{c}
\mu  {\bf 1}_{\widetilde{N}_c''}  \\
0
\end{array}
\right), 
\qquad
<\widetilde{g}>   = 
\left(
\begin{array}{cc}
\mu  {\bf 1}_{\widetilde{N}_c''} & 0  \\
\end{array}
\right), 
\qquad
<\Phi'> =
 \left(
\begin{array}{cc}
0  & 0  \\
0 & \Phi_0'  {\bf 1}_{(N_c'-N_c)-\widetilde{N}_c''} 
\end{array}
\right) 
\nonu
\eea
where the zero of $<g>$ is a $
(N_c'-N_c-\widetilde{N}_c'') \times \widetilde{N}_c''$ 
matrix, the zero of $<\widetilde{g}>$ is a
$\widetilde{N}_c'' \times (N_c'-N_c-\widetilde{N}_c'') $ matrix and 
the zeros of $<\Phi'>$ are $\widetilde{N}_c'' \times \widetilde{N}_c''$,
$\widetilde{N}_c'' \times 
(N_c'-N_c-\widetilde{N}_c'')$ and $(N_c'-N_c-\widetilde{N}_c'') \times
\widetilde{N}_c''$ 
matrices.
Then one can expand these fields around on a point, as
in \cite{ISS,Shih} and one arrives at the relevant superpotential
up to quadratic order in the fluctuation. 
At one loop, the effective potential $V_{eff}^{(1)}$ for $\Phi'_0$
leads to the positive value for $m_{\Phi'_0}^2$ implying that these
vacua are stable.

\section{Meta-stable brane configurations with four NS-branes plus O4-plane}

In this section, we add an orientifold 4-plane to the previous brane
configurations and find out new meta-stable brane configurations.
Or one can realize these brane configurations by inserting the  
extra NS-brane and O4-planes into the brane configuration \cite{Ahn07-2}.
 
\subsection{Electric theory}

The type IIA brane configuration  \cite{Ahn97} corresponding to 
${\cal N}=1$ supersymmetric gauge theory with
gauge group
\bea
Sp(N_c) \times SO(2N_c') \times Sp(N_c'')
\label{elec}
\eea
and with a field $F$ charged under
$({\bf 2N_c, 2N_c'})$, a field $G$ charged under
$({\bf 2N_c', 2N_c''})$ 
can be described by 
the left $NS5_L'$-brane, the left 
$NS5_L$-brane,
the right $NS5_R'$-brane, the right 
$NS5_R$-brane,
 $2N_c$-, $2N_c'$-  and $2N_c''$-color D4-branes 
 as
well as an 
$O4^{\pm}$-plane(01236) we should add. 
The $O4^{\pm}$-planes act as $(x^4,x^5,x^7,x^8,x^9) \rightarrow
(-x^4,-x^5,-x^7,
-x^8,-x^9)$ as usual 
and they have RR charge $\pm 1$ playing the role of $\pm 1$
D4-brane.

We draw this brane configuration in Figure 6A for the vanishing mass
case. 

\begin{figure}[ht]
   \epsfxsize=5.0in 
\centerline{\epsffile{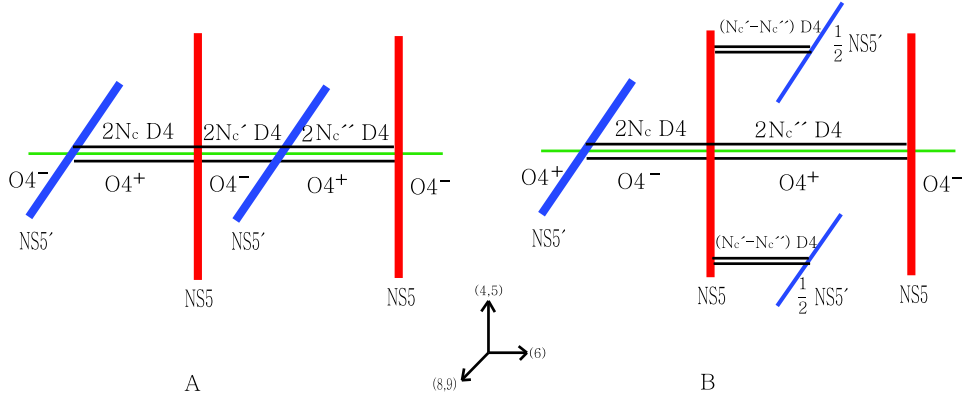}}
   \caption[FIG. \arabic{figure}.]{ 
The  
  ${\cal N}=1$ supersymmetric 
electric brane configuration for the gauge group $Sp(N_c) \times
SO(2N_c') \times Sp(N_c'')$ 
and bifundamentals $F$ and $G$  with vanishing(6A) and 
nonvanishing(6B) mass
for the bifundamental $F$. 
The $2N_c'$ D4-branes in 6A are decomposed into 
$2(N_c'-N_c'')$ D4-branes which are moving to $\pm v$ direction in
  ${\bf Z}_2$ symmetric way in 6B 
and $2N_c''$ D4-branes which are recombined with those D4-branes
connecting between $NS5_R'$-brane and $NS5_R$-brane in 6B.
  }
\end{figure}

There is no superpotential in Figure 6A. Let us deform this theory.
Displacing the two NS5'-branes relative each other in the $+v$ 
direction corresponds to turning on a quadratic
mass-deformed superpotential
for the field $F$ as follows:
\bea
W = m F F \equiv m \Phi'
\label{mass1}
\eea
where 
the first gauge group indices in $F$  
are contracted, each second gauge group index in $F$ is encoded in 
$\Phi'$
 and the mass $m$ is given by (\ref{m}).
The gauge-singlet $\Phi'$ for the first dual gauge group is in the 
adjoint representation for the second dual gauge group, 
i.e., ${(\bf 1, (N_c'-N_c'')(2N_c'-2N_c''-1),1) }$ 
under the dual gauge group (\ref{Dual}). 
Then the $\Phi'$ is a $2(N_c'-N_c'') \times 2(N_c'-N_c'')$ matrix.
The half $NS5_R'$-brane \cite{BFH} 
together with $(N_c'-N_c'')$-color D4-branes 
is moving to the $+v$ direction(and their mirrors to $-v$ direction) for
fixed other branes during this mass deformation. 
The $2N_c''$ D4-branes among $2N_c'$ D4-branes 
are not participating in 
the mass deformation.
Then the $x^5$ coordinate 
of $NS5_L'$-brane is equal to
zero
while the $x^5$ coordinates of half $NS5_R'$-brane are given by 
$\pm \Delta x$.

Giving an expectation value to the meson field $\Phi'$
corresponds to recombination of $2N_c$- and $2N_c'$- color 
D4-branes, which will become $2N_c$-color D4-branes
in Figure 6A such that they are suspended between 
the $NS5_L'$-brane and the $NS5_R'$-brane 
and pushing them into the $w$
direction. We assume that the number of colors satisfies
\bea
N_c' \geq N_c +2 \geq N_c''.
\nonu
\eea

Now 
we draw this brane configuration in Figure 6B for nonvanishing mass
for the field $F$. 

\subsection{Magnetic theory}

By applying the Seiberg dual to the $Sp(N_c)$ factor in 
(\ref{elec}), the two $NS5_{L,R}'$-branes can be located at the
inside of the two NS5-branes, as in Figure 7.
Starting from Figure 6B and interchanging the 
$NS5_L'$-brane and the $NS5_L$-brane,
one obtains the Figure 7A.

\begin{figure}[ht]
   \epsfxsize=5.0in 
\centerline{\epsffile{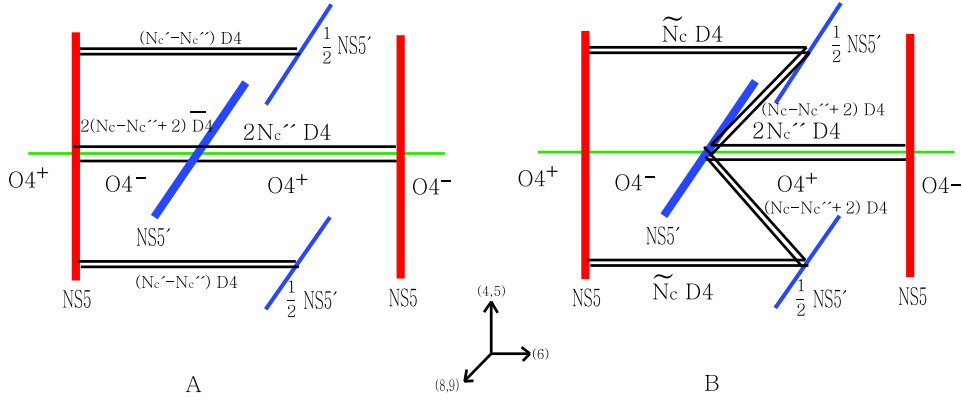}}
   \caption[FIG. \arabic{figure}.]{ 
The  ${\cal N}=1$ magnetic brane configuration for the gauge group 
$Sp(\widetilde{N}_c=N_c'-N_c-2) 
\times SO(2N_c') \times Sp(N_c'')$ corresponding to Figure 6B with D4-
and $\overline{D4}$-branes(7A) and with 
a misalignment between D4-branes(7B) when the NS5'-branes are close to
each other.
 The number of tilted D4-branes in 7B can be written as
$N_c-N_c''+2=(N_c'-N_c'')-\widetilde{N}_c$.  }
\end{figure}

Before arriving at the Figure 7A, there exists an intermediate 
step where the $2(N_c'-N_c-2)$ D4-branes are connecting between the 
$NS5_L$-brane and the  $NS5_L'$-brane,  
$(N_c'-N_c'')$ D4-branes connecting between the  $NS5_L'$-brane and   
$NS5_R'$-brane(and their mirrors), 
and $2N_c''$ D4-branes between the $NS5_L'$-brane and
the $NS5_R$-brane. By introducing $-2N_c''$ D4-branes and $-2N_c''$ 
anti-D4-branes  between the  $NS5_L$-brane and   
$NS5_L'$-brane, reconnecting the former with  
the $N_c'$ D4-branes connecting between  
$NS5_L$-brane
and the $NS5_L'$-brane  (therefore $(N_c'-N_c'')$ D4-branes)
and moving those combined
$(N_c'-N_c'')$ 
D4-branes
to $+v$-direction(and their mirrors to $-v$ direction), 
one gets the final Figure 7A where we are left with 
$2(N_c-N_c''+2)$ 
anti-D4-branes between the $NS5_L$-brane and   
$NS5_L'$-brane.
When two NS5'-branes in Figure 7A are close to each other, then 
it leads to Figure 7B
 by realizing that the number of $(N_c'-N_c'')$
D4-branes connecting between $NS5_L$-brane and $NS5_R'$-brane can
be rewritten as $(N_c-N_c''+2)$ plus $\widetilde{N}_c$.

The dual gauge group is 
\bea
Sp(\widetilde{N}_c=N_c'-N_c-2) \times SO(2N_c') \times Sp(N_c'')
\label{Dual}.
\eea
The matter contents are the field $f$ 
 charged under
$({\bf 2\widetilde{N}_c, 2N_c', 1})$, a field $g$ charged under
$({\bf 1, 2N_c', 2N_c''})$ under the dual gauge group
(\ref{Dual})
and  
the gauge-singlet $\Phi'$ that is in the 
adjoint representation 
for the second dual gauge group, 
i.e., 
 ${(\bf 1, (N_c'-N_c'')(2N_c'-2N_c''-1),1) }$ under the 
dual gauge group. That is, the $\Phi'$ is  an 
 $2(N_c'-N_c'') \times 2(N_c'-N_c'')$ antisymmetric matrix.

The cubic superpotential with the mass term (\ref{mass1}) in the dual
theory is given by
\bea
W_{dual} = \Phi' f f + m \Phi'. 
\label{sup}
\eea
Here the magnetic field $f$  
corresponds to 4-4 strings connecting 
the $2\widetilde{N}_c$-color D4-branes(that are 
connecting between the $NS5_L$-brane
and the $NS5_R'$-brane including the mirrors) with $2N_c'$-flavor 
D4-branes(that is 
a combination of three different D4-branes including the mirrors 
in Figure 7B).
Among these $2N_c'$-flavor D4-branes, only the strings ending on
the upper $2(N_c'-N_c-2)$ D4-branes and 
on the tilted middle $2(N_c-N_c''+2)$ 
D4-branes including the mirrors 
in Figure 7B enter the cubic superpotential term. 
Although the $(N_c'-N_c'')$ D4-branes in Figure 7A cannot move any
directions,
the tilted $2(N_c-N_c''+2)$-flavor D4-branes including the mirrors 
can move $w$ direction.
The remaining upper $\widetilde{N}_c$ D4-branes(and its mirrors) 
are fixed also and cannot 
move any direction. 
Note that 
there is a decomposition 
\bea
(N_c'-N_c'')=(N_c-N_c''+2)+\widetilde{N}_c.
\nonu
\eea 

The brane configuration for zero mass for the bifundamental,
which has only a cubic superpotential,
can be obtained from Figure 7A by moving
the upper and lower NS5'-branes together with $(N_c'-N_c'')$ color D4-branes 
into the origin $v=0$.
Then the number of dual colors for D4-branes 
becomes  
$2\widetilde{N}_c$ between $NS5_L$-brane and $NS5_L'$-brane
and $2N_c'$ between two NS5'-branes as well as $2N_c''$ between
$NS5_R'$-brane and $NS5_R$-brane.
Or starting from Figure 6A and moving the $NS5_L$-brane to the left all the
way past the $NS5_L'$-brane,
one also obtains the corresponding magnetic brane configuration
for massless case.

The brane configuration in Figure 7A is stable as long as the
distance $\Delta x$ between the upper NS5'-brane and 
the middle NS5'-brane is large. If they are close to each other, then this brane
configuration is unstable to decay and leads to 
the brane configuration in Figure
7B.
One can regard these brane configurations as particular states in the
magnetic gauge theory with the gauge group (\ref{Dual}) and
superpotential (\ref{sup}).
The  upper $(N_c'-N_c''-\widetilde{N}_c)$ flavor D4-branes of 
straight brane configuration
of
Figure 7B  bend due to the fact that there exists an attractive
gravitational interaction
between those flavor D4-branes and $NS5_L$-brane from the DBI action, 
as long as the distance $y_3$
goes to $\infty$ because the presence of an extra $NS5_R$-brane does
not affect the DBI action. 
For the finite and small $y_3$, the careful analysis for DBI action is
needed in
order to obtain the bending curve connecting  two NS5'-branes.  
Of course, their mirrors, the lower 
$(N_c'-N_c''-\widetilde{N}_c')$ flavor D4-branes of 
straight brane configuration
of
Figure 7B can bend and their trajectories connecting 
two NS5'-branes should be preserved under the O4-plane, i.e., ${\bf
Z}_2$ symmetric way.

When the upper and lower half $NS5_R'$-branes 
are replaced by coincident $(N_c'-N_c'')$ 
D6-branes and 
the $NS5_R$ is rotated by an angle $\frac{\pi}{2}$ in the $(v,w)$
plane in Figure 7B, this brane configuration reduces to the one 
found in \cite{Ahn07-2} where the gauge group was given by 
$Sp(n_f+n_c'-n_c-2) \times SO(2n_c')$ 
with $2n_f$ multiplets, flavor singlet and gauge singlets. 
Then the present $(N_c'-N_c'')$ corresponds to the $n_f$,
the number $N_c$ corresponds to $n_c$ and 
$N_c''$ corresponds to the $n_c'$ of \cite{Ahn07-2}. 
However, the gauge group $Sp(N_c'')$ corresponds to the different 
gauge group $SO(2n_c')$. When we discuss the subsection 3.5 and take
the Seiberg dual for the middle gauge group, then it becomes
$SO(2N_c) \times Sp(\widetilde{N}_c'=N_c+N_c''-N_c'-2) \times
SO(2N_c'')$.
Then the $N_c$ corresponds to the $n_f$,
the number $N_c'$ corresponds to $n_c$ and 
$N_c''$ corresponds to the $n_c'$ of \cite{Ahn07-2}. 
If we ignore $2N_c''$ D4-branes and $NS5_R$-brane from Figure 7B, then
the brane configuration becomes the one in \cite{Ahn06-1,FGU}.

The dual gauge theory has  an adjoint $\Phi'$ of $SO(2N_c')$ and 
bifundamentals $f$ and $g$  under the dual gauge
group (\ref{Dual}) and the superpotential 
corresponding to Figures 7A and 7B is given by 
\bea
W_{dual} = h \Phi' f f - h \mu^2 \Phi', \qquad h^2 = g_{2,
  mag}^2,
\qquad \mu^2 = -\frac{\Delta x}{ 2\pi g_s \ell_s^3}.
\nonu
\eea
Then $ f f$ is a $2\widetilde{N}_c \times 2\widetilde{N}_c$ 
matrix where the second gauge group indices for $f$  
are contracted with those
of $\Phi'$ while $\mu^2$ is a 
$2(N_c'-N_c'') \times 2(N_c'-N_c'')$ matrix.
The product $f f$ has the same representation for the 
product of quarks
and moreover, 
the first gauge group indices for the field $\Phi'$ play the
role of the flavor indices, as we observed above for the comparison
with the brane configuration in the presence of D6-branes.

Therefore, the F-term equation, the derivative $W_{dual}$ with respect to the
meson field $\Phi'$ cannot be satisfied if the $2(N_c'-N_c'')$ exceeds
$2\widetilde{N}_c$.
So the supersymmetry is broken.   
That is, 
there exist two equations from F-term conditions:
$
f f -\mu^2 =0$ and $ \Phi' f =0$.
Then the solutions for these
are given by 
\bea
<f>   = 
\left(
\begin{array}{c}
\mu  {\bf 1}_{2\widetilde{N}_c}  \\
0
\end{array}
\right), 
\qquad
<\Phi'> =
 \left(
\begin{array}{cc}
0  & 0  \\
0 & \Phi'_0  {\bf 1}_{(N_c'-N_c''-\widetilde{N}_c) \otimes i \sigma_2} 
\end{array}
\right) 
\label{poi}
\eea
where the zero of $<f>$ is a $
2(N_c'-N_c''-\widetilde{N}_c) \times 2\widetilde{N}_c$ 
matrix and 
the zeros of $<\Phi'>$ are $2\widetilde{N}_c \times 2\widetilde{N}_c$,
$2\widetilde{N}_c \times 2(N_c'-N_c''-\widetilde{N}_c)$, 
and $2(N_c'-N_c''-\widetilde{N}_c) \times
2\widetilde{N}_c$ matrices.
Then one can expand these fields around on a point (\ref{poi}), as
in \cite{ISS} and one arrives at the relevant superpotential
up to quadratic order in the fluctuation. 
At one loop, the effective potential $V_{eff}^{(1)}$ for $\Phi'_0$
leads to the positive value for $m_{\Phi'_0}^2$ implying that these
vacua are stable.

\subsection{Other magnetic theory-I}

Let us consider other magnetic theory for the same electric theory
given in the subsection 3.1.
By applying the Seiberg dual to the $SO(2N_c')$ factor in 
(\ref{elec}), the two $NS5_{L,R}'$-branes can be located at the
left hand side of the two NS5-branes, as in Figure 9.

\begin{figure}[ht]
   \epsfxsize=5.0in 
\centerline{\epsffile{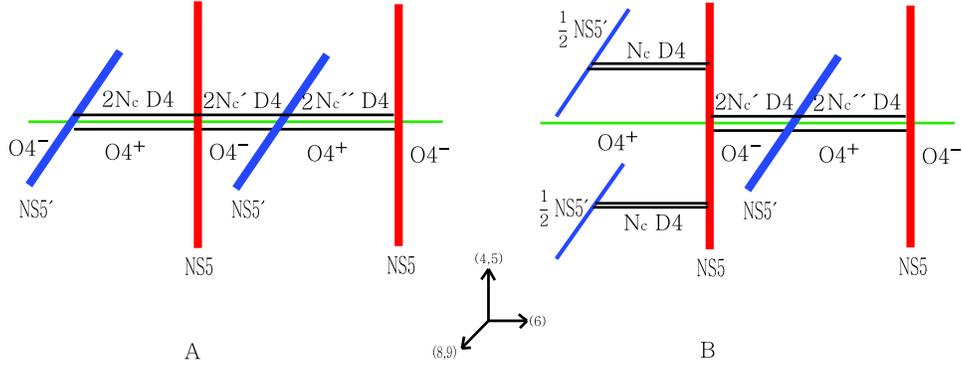}}
   \caption[FIG. \arabic{figure}.]{ 
The  
  ${\cal N}=1$ supersymmetric 
electric brane configuration for the gauge group $Sp(N_c) \times
SO(2N_c') \times Sp(N_c'')$ 
and  bifundamentals $F$ and $G$  with vanishing(8A) which is the same
  as Figure 6A and 
nonvanishing(8B) mass
for the bifundamental $F$. This deformation is different from the
  previous case (\ref{mass1}). 
 In 8B, the $2N_c$ D4-branes are moving to $\pm v$ directions in 
${\bf Z}_2$ symmetric way.  }
\end{figure}

The Figure 8A is the same as the one in Figure 6A and one moves 
half $NS5_L'$-brane together with $N_c$ D4-branes to $+v$ 
direction(and its mirrors to $-v$ direction)  and 
is given by Figure 8B.
Starting from Figure 8B and interchanging the $NS5_L$-brane 
and the $NS5_R'$-brane,
one obtains the Figure 9A.
Before arriving at the Figure 9A, there exists an intermediate 
step where the $N_c$ D4-branes are connecting between the 
$NS5_L'$-brane and the $NS5_R'$-brane(and their mirrors),  
$2(N_c''-N_c'+N_c+2)$ D4-branes are connecting between the $NS5_R'$-brane and   
$NS5_L$-brane, and $2N_c''$ D4-branes are suspended 
between the $NS5_L$-brane and
the $NS5_R$-brane. 
By moving the combined
$N_c$ D4-branes, obtained from the reconnection of those
D4-branes between the $NS5_L'$-brane and the $NS5_R'$-brane and those
D4-branes between the $NS5_R'$-brane and the $NS5_L$-brane(therefore 
between the $NS5_L'$-brane and the $NS5_L$-brane), 
to $+v$-direction(and their mirrors to $-v$ direction), 
one gets the final Figure 9A where we are left with 
$2(N_c'-N_c''-2)$ 
anti-D4-branes between the $NS5_R'$-brane and   
$NS5_L$-brane.
We assume  that the number of colors satisfies
\bea
N_c+N_c'' \geq N_c'-2 \geq N_c''.
\nonu
\eea
When two NS5'-branes in Figure 9A are close to each other, then 
it leads to Figure 9B
 by realizing that the number of $N_c$
D4-branes connecting between $NS5_L'$-brane and $NS5_L$-brane in
Figure 9A can
be rewritten as $(N_c'-N_c''-2)$ plus $\widetilde{N}_c'$.
If we ignore $2N_c''$ D4-branes and $NS5_R$-brane and change the
O4-plane charge(corresponding to change the symplectic gauge group into the
orthogonal gauge group and vice versa) from Figure 9B, then
the brane configuration becomes the one in \cite{Ahn07-5}.

\begin{figure}[ht]
   \epsfxsize=5.0in 
\centerline{\epsffile{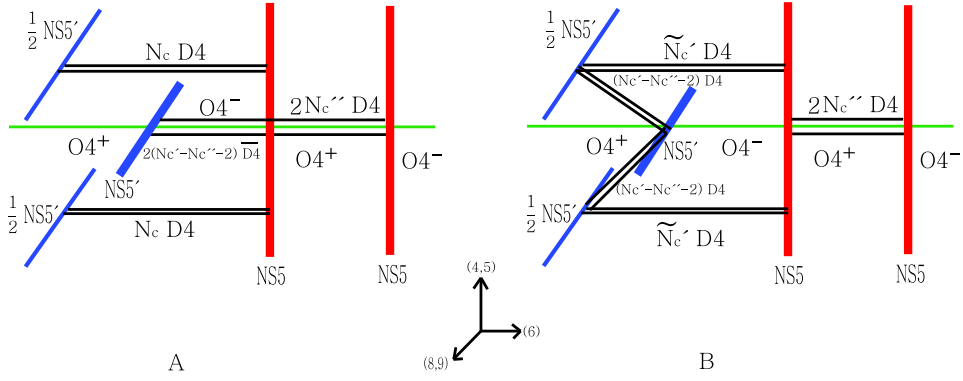}}
   \caption[FIG. \arabic{figure}.]{ 
The  ${\cal N}=1$ 
magnetic brane configuration for the gauge group $Sp(N_c) \times 
SO(2\widetilde{N}_c'=2N_c+2N_c''-2N_c'+4) \times
Sp(N_c'')$ 
corresponding to Figure 8B with D4-
and $\overline{D4}$-branes(9A) and with 
a misalignment between D4-branes(9B) when the NS5'-branes are close to
each other.
 The number of tilted D4-branes is equal to $N_c'-N_c''-2=
N_c-\widetilde{N}_c'$ in 9B.  }
\end{figure}

The dual gauge group is 
\bea
Sp(N_c) \times SO(2\widetilde{N}_c'=2N_c+2N_c''-2N_c'+4) \times Sp(N_c'').
\label{dualdual}
\eea
The matter contents are the field $f$ 
 charged under
$({\bf 2N_c, 2\widetilde{N}_c', 1})$, a field $g$ charged under
$({\bf 1, 2\widetilde{N}_c', 2N_c''})$ under the dual gauge group
(\ref{dualdual})
and  
the gauge-singlet $\Phi$ for the second dual gauge group in the 
adjoint representation for the first dual gauge group, 
i.e., ${(\bf N_c(2N_c+1),1,1) }$ under the 
dual gauge group.
Then the $\Phi$ is a $2N_c \times 2N_c$ matrix.
All the $2N_c$ D4-branes are participating in the mass deformation.

The cubic superpotential with the mass term  in the dual
theory 
\footnote{One can also construct the mass deformation by
  rotating $NS5_R$-brane and  moving
it to $\pm v$ direction, as in previous case in subsection 2.3. 
The brane configuration 
can be obtained easily.  }
is given by
\bea
W_{dual} = \Phi f f + m \Phi 
\label{super}
\eea
where we define $\Phi$ as $\Phi \equiv F F$ and 
the second gauge group indices in $F$ 
are contracted, each first gauge group index in them is encoded in 
$\Phi$. Although the $\Phi$ that has first gauge group indices 
looks similar to the previous 
$\Phi'$ that has second gauge group indices, 
the group indices are different.
Here the magnetic field $f$ 
corresponds to 4-4 strings connecting 
the $2\widetilde{N}_c'$-color D4-branes(that are 
connecting between the $NS5_L'$-brane
and the $NS5_L$-brane in Figure 9B) with $2N_c$-flavor 
D4-branes including the mirrors(which  are realized 
as corresponding D4-branes in Figure 9A).
Although the $N_c$ D4-branes(and its mirrors) in Figure 9A cannot move any
directions,
the tilted $(N_c'-N_c''-2)$-flavor D4-branes(and its mirrors) can move 
$w$ direction in Figure 9B.
The remaining upper $\widetilde{N}_c'$ D4-branes(and its mirrors) 
are fixed also and cannot 
move any direction. 
Note that 
there is a decomposition 
\bea
N_c=(N_c'-N_c''-2)+\widetilde{N}_c'.
\nonu
\eea

The brane configuration for zero mass for the bifundamental,
which has only a cubic superpotential,
can be obtained from Figure 9A by moving
the upper and lower NS5'-branes together with $N_c$ color D4-branes 
into the origin $v=0$.
Then the number of dual colors for D4-branes 
becomes $2N_c$ between two NS5'-branes, 
 $2\widetilde{N}_c'$ between $NS5_R'$-brane and $NS5_L$-brane
and $2N_c''$ 
between $NS5_L$-brane and $NS5_R$-brane.
Or starting from Figure 8A and moving the $NS5_L$-brane to the right all the
way past the $NS5_R'$-brane,
one also obtains the corresponding magnetic brane configuration
for massless case.

The brane configuration in Figure 9A is stable as long as the
distance $\Delta x$ between the upper NS5'-brane and 
the middle NS5'-brane is large. If they are close to each other, then this brane
configuration is unstable to decay and leads to 
the brane configuration in Figure
9B.
One can regard these brane configurations as particular states in the
magnetic gauge theory with the gauge group (\ref{dualdual}) and
superpotential (\ref{super}).
The  $(N_c-\widetilde{N}_c')$ flavor D4-branes of 
straight brane configuration
of
Figure 9B  bend due to the fact that there exists an attractive
gravitational interaction
between those flavor D4-branes and $NS5_L$-brane from the DBI action, 
as long as the distance $y_3$
goes to $\infty$ because the presence of an extra $NS5_R$-brane does
not affect the DBI action. 
For the finite and small $y_3$, the careful analysis for DBI action is
needed in
order to obtain the bending curve connecting  two NS5'-branes.  
Or if $y_3$ goes to zero, then this extra $NS5_R$-brane plays the role
of enhancing the strength for the  
NS5-branes and will affect both the energy of bending curve, 
$E_{curved}$, and $\Delta x$ \cite{GK}.
Of course, their mirrors, the lower 
$(N_c-\widetilde{N}_c')$ flavor D4-branes of 
straight brane configuration
of
Figure 9B can bend and their trajectories connecting 
two NS5'-branes should be preserved under the O4-plane, i.e., ${\bf
Z}_2$ symmetric way.

When the NS5'-brane(or $NS5_L'$-brane) 
is replaced by coincident $N_c$ 
D6-branes, this brane configuration looks similar to the one 
found in \cite{Ahn07-2} where the gauge group was given by 
$SO(2n_f'+2n_c-2n_c'+4) \times Sp(n_c)$ 
with $2n_f'$ multiplets, flavor singlet and gauge singlets. 
Then the present $N_c$ corresponds to the $n_f'$, the number $N_c'$
corresponds to $n_c'$ and 
$N_c''$ corresponds to the $n_c$ of \cite{Ahn07-2}. 

The dual gauge theory has  an adjoint $\Phi$ of $Sp(N_c)$ and 
bifundamentals $f$ and $ g$  under the dual gauge
group (\ref{dualdual}) and the superpotential 
corresponding to Figures 9A and 9B is given by 
\bea
W_{dual} = h \Phi f f - h \mu^2 \Phi, \qquad h^2 = g_{1,
  mag}^2,
\qquad \mu^2 = -\frac{\Delta x}{ 2\pi g_s \ell_s^3}.
\nonu
\eea
Then $ f f$ is a $2\widetilde{N}_c' \times 2\widetilde{N}_c'$ 
matrix where the first gauge group indices for $f$  
are contracted with those
of $\Phi$ while $\mu^2$ is a 
$2N_c \times 2N_c$ matrix.
The product $f f$ has the same representation for the 
product of quarks
and moreover, 
the first gauge group indices for the field $\Phi$ play the
role of the flavor indices as we observed above.

Therefore, the F-term equation, the derivative $W_{dual}$ with respect to the
meson field $\Phi$ cannot be satisfied if the $2N_c$ exceeds
$2\widetilde{N}_c'$.
So the supersymmetry is broken.   
That is, 
there exist two equations from F-term conditions:
$
f f -\mu^2 =0$ and $ \Phi f =0$.
Then the solutions for these
are given by 
\bea
<f>   = 
\left(
\begin{array}{c}
\mu  {\bf 1}_{2\widetilde{N}_c'}  \\
0
\end{array}
\right), 
\qquad
<\Phi> =
 \left(
\begin{array}{cc}
0  & 0  \\
0 & \Phi_0  {\bf 1}_{2(N_c-\widetilde{N}_c')} 
\end{array}
\right) 
\label{point2}
\eea
where the zero of $<f>$ is a $
2(N_c-\widetilde{N}_c') \times 2\widetilde{N}_c'$ 
matrix and 
the zeros of $<\Phi>$ are $2\widetilde{N}_c' \times 2\widetilde{N}_c'$,
$2\widetilde{N}_c' \times 
2(N_c-\widetilde{N}_c')$ and $2(N_c-\widetilde{N}_c') \times
2\widetilde{N}_c'$ 
matrices.
Then one can expand these fields around on a point (\ref{point2}), as
in \cite{ISS} and one arrives at the relevant superpotential
up to quadratic order in the fluctuation. 
At one loop, the effective potential $V_{eff}^{(1)}$ for $\Phi_0$
leads to the positive value for $m_{\Phi_0}^2$ implying that these
vacua are stable.

\subsection{Other magnetic theory-II}

One can think of the following dual gauge group
\bea
Sp(N_c) \times SO(2N_c') \times Sp(\widetilde{N}_c''=N_c'-N_c''-2)
\label{dualdualdual}
\eea
by performing the magnetic dual for the last gauge group in (\ref{elec}).
The electric brane configuration can be given in terms of Figure 6A or
Figure 6A with an exchange between NS5-brane and NS5'-brane. For the
latter, 
the
resulting brane configuration is given by $NS5_L$-brane,
$NS5_L'$-brane, $NS5_R$-brane, and $NS5'_R$-brane from the left to the
right in the $x^6$ direction.  One can take the magnetic dual 
either by following
the previous procedure or by looking at the Figure 7 from the negative
$w$ direction which is an opposite viewpoint, compared with Figure 7.
In other words, we are looking at the Figure 7 from the other side of
$w$.

Then the resulting brane configuration in this case can be obtained by 
taking a reflection for all the NS-branes, D4-branes and anti
D4-branes with respect to the $NS5_L$-brane(rotating them to the left
for fixed $NS5_L$-brane) in Figure 7A and Figure 7B. 
Then  the ${\cal N}=1$ magnetic brane configuration for the gauge group 
$Sp(N_c) 
\times SO(2N_c') \times Sp(\widetilde{N}_c''=N_c'-N_c''-2)$ 
corresponds to  the Figure 10A' with D4-
and $\overline{D4}$-branes and the Figure 10B' with 
a misalignment between D4-branes when the NS5'-branes are close to
each other. The number of tilted D4-branes in 10B' can be written as
$N_c''-N_c+2=(N_c'-N_c)-\widetilde{N}_c''$. 
We do not present the Figures 10A' and 10B' here.

We turn to the other case.
Let us consider other magnetic theory for the same electric theory
given in the subsection 3.1 with Figure 6A.
By applying the Seiberg dual to the $Sp(N_c'')$ factor in 
(\ref{elec}) from Figure 6A and 
interchanging the $NS5_R'$-brane and the $NS5_R$-brane,
one obtains the Figure 10A''.
Before arriving at the Figure 10A'', there exists an intermediate 
step where 
$2N_c$ D4-branes between
$NS5_L'$-brane and the $NS5_L$-brane,
the $2N_c'$ D4-branes are connecting between the 
$NS5_L$-brane and the  $NS5_R$-brane,  
$(N_c'-N_c''-2)$ D4-branes are connecting between the  $NS5_R$-brane and   
$NS5_R'$-brane(and their mirrors).
By rotating $NS5_L$-brane by an angle $\frac{\pi}{2}$, 
moving it with the $(N_c'-N_c)$ D4-branes 
to $+v$ direction where we introduce $2(N_c'-N_c)$ D4-branes and
$2(N_c'-N_c)$ anti D4-branes between the $NS5_R$-brane and the 
$NS5_R'$-brane, 
one gets the final Figure 10A'' where we are left with 
$2(N_c''-N_c+2)$ 
anti-D4-branes between the $NS5_R$-brane and   
the $NS5_R'$-brane.
When two NS5'-branes in Figure 10A'' are close to each other, then 
it leads to Figure 10B''
 by realizing that the number of $(N_c'-N_c)$
D4-branes connecting between $NS5_M'$-brane and NS5-brane can
be rewritten as $(N_c''-N_c+2)$ plus $\widetilde{N}_c''$.

The brane configuration in Figure 10A'' is stable as long as the
distance $\Delta x$ between the upper NS5'-brane and 
the middle NS5'-brane(or $NS5_R'$-brane) 
is large. If they are close to each other, then this brane
configuration is unstable to decay to 
the brane configuration in Figure
10B''.
One can regard these brane configurations as particular states in the
magnetic gauge theory with the gauge group and
superpotential.
The  upper $(N_c'-N_c-\widetilde{N}_c'')$ flavor D4-branes of 
straight brane configuration
of
Figure 10B''  bend since there exists an attractive
gravitational interaction
between those flavor D4-branes and NS5-brane from the DBI action. 
As mentioned in \cite{Ahn07-5},
the two NS5'-branes are located at different side of NS5-brane in
Figure 10B'' and the DBI action computation for this bending curve
should be taken into account. 
Of course, their mirrors, the lower 
$(N_c'-N_c-\widetilde{N}_c'')$ flavor D4-branes of 
straight brane configuration
of
Figure 10B'' can bend and their trajectories connecting 
two NS5'-branes should be preserved under the O4-plane, i.e., ${\bf
Z}_2$ symmetric way.

\begin{figure}[ht]
   \epsfxsize=5.0in 
\centerline{\epsffile{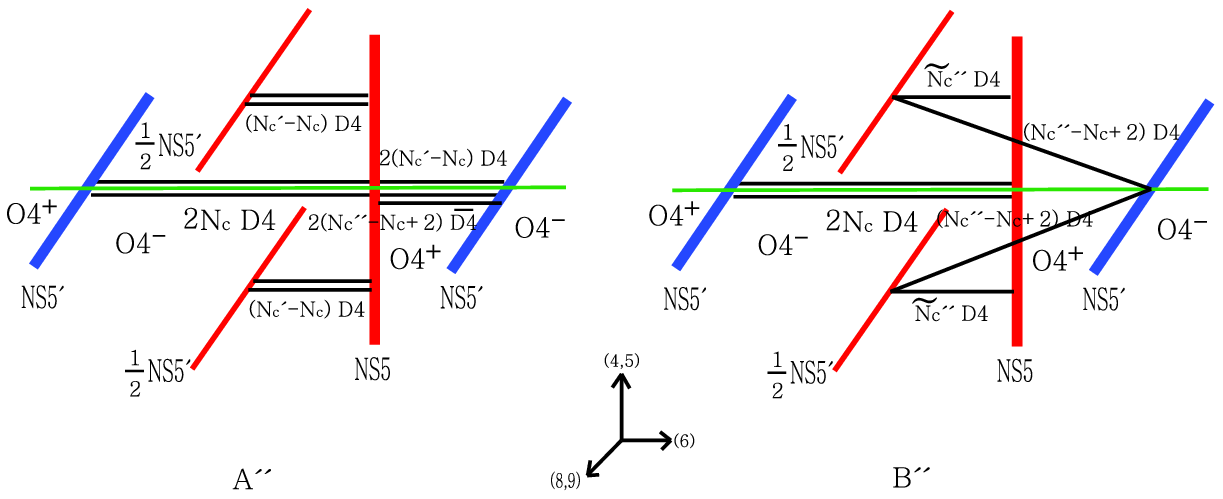}}
   \caption[FIG. \arabic{figure}.]{ 
The  ${\cal N}=1$ magnetic brane configuration for the gauge group 
$Sp(N_c) 
\times SO(2N_c') \times Sp(\widetilde{N}_c''=N_c'-N_c''-2)$ 
with D4-
and $\overline{D4}$-branes(10A'') and with 
a misalignment between D4-branes(10B'') when the NS5'-branes are close to
each other.
The number of tilted D4-branes in 10B'' can be written as
$N_c''-N_c+2=(N_c'-N_c)-\widetilde{N}_c''$. 
The deformation is rleated to the bifundamentals $G$.  }
\end{figure}

The matter contents are the field $f$ 
 charged under
$({\bf 2N_c, 2N_c', 1})$, a field $g$ charged under
$({\bf 1, 2N_c', 2\widetilde{N}_c''})$ under the dual gauge group
(\ref{dualdualdual})
and  
the gauge-singlet $\Phi'$ that is in the 
adjoint representation 
for the second dual gauge group, 
i.e., 
 ${(\bf 1, (N_c'-N_c)(2N_c'-2N_c-1),1) }$ under the 
dual gauge group. That is, the $\Phi'$ is  an 
 $2(N_c'-N_c) \times 2(N_c'-N_c)$ antisymmetric matrix.

The cubic superpotential with the mass term  in the dual
theory is given by
\bea
W_{dual} = \Phi' g g + m \Phi' 
\label{sup-1}
\eea
where we define $\Phi'$ as $\Phi' \equiv G G$ and 
the third gauge group indices in $G$  
are contracted, each seond gauge group index in $G$ is encoded in 
$\Phi'$. Although the $\Phi'$ that has second gauge group indices 
looks similar to the previous 
$\Phi$ that has first gauge group indices, 
the group indices are different.
Here the magnetic field $g$  
correspond to 4-4 strings connecting 
the $2\widetilde{N}_c''$-color D4-branes including the mirrors(that are 
connecting between the $NS5_M'$-brane
and the NS5-brane in Figure 10B'') with $2N_c'$-flavor 
D4-branes.
Among these $2N_c'$-flavor D4-branes, only the strings ending on
the upper $2(N_c'-N_c''-2)$ D4-branes and 
on the tilted middle $2(N_c''-N_c+2)$ 
D4-branes in Figure 10B'' enter the cubic superpotential term. 
Although the $(N_c'-N_c)$ D4-branes(and its mirrors) 
in Figure 10A'' cannot move any
directions,
the tilted $(N_c''-N_c+2)$-flavor D4-branes(and its mirrors) 
can move $w$ direction.
The remaining upper and lower 
$\widetilde{N}_c''$ D4-branes are fixed also and cannot 
move any direction. 
Note that 
there is a decomposition 
\bea
(N_c'-N_c)=(N_c''-N_c+2)+\widetilde{N}_c''.
\nonu
\eea 

The brane configuration for zero mass for the bifundamental,
which has only a cubic superpotential,
can be obtained from Figure 10A'' by moving
the upper and lower NS5'-branes together with $(N_c'-N_c)$ color D4-branes 
into the origin $v=0$.
Then the number of dual colors for D4-branes 
becomes $2N_c$ between two NS5'-branes 
and $2N_c'$ between the $NS5_M'$-brane and the NS5-brane and 
$2\widetilde{N}_c''$ between the NS5-brane and the $NS5_R'$-brane.
Or starting from Figure 6A and moving the $NS5_R$-brane to the left all the
way past the $NS5_R'$-brane,
one also obtains the corresponding magnetic brane configuration
for massless case.

The dual gauge theory has  an adjoint $\Phi'$ of $SO(2N_c')$ and 
bifundamentals $f$ and $g$  under the dual gauge
group (\ref{dualdualdual}) and the superpotential 
corresponding to Figures 10A'' and 10B'' is given by 
\bea
W_{dual} = h \Phi' g g - h \mu^2 \Phi', \qquad h^2 = g_{2,
  mag}^2,
\qquad \mu^2 = -\frac{\Delta x}{ 2\pi g_s \ell_s^3}.
\nonu
\eea
Then $ g g$ is a $2\widetilde{N}_c'' \times 2\widetilde{N}_c''$ 
matrix where the second gauge group indices for $g$  
are contracted with those
of $\Phi'$ while $\mu^2$ is a 
$2(N_c'-N_c) \times 2(N_c'-N_c)$ matrix.
The product $g g$ has the same representation for the 
product of quarks
and moreover, 
the first gauge group indices for the field $\Phi'$ play the
role of the flavor indices.

Therefore, the F-term equation, the derivative $W_{dual}$ with respect to the
meson field $\Phi'$ cannot be satisfied if the $2(N_c'-N_c)$ exceeds
$2\widetilde{N}_c''$.
So the supersymmetry is broken.   
That is, 
there exist two equations from F-term conditions:
$
g g -\mu^2 =0$ and $ \Phi' g =0$.
Then the solutions for these
are given by 
\bea
<g>   = 
\left(
\begin{array}{c}
\mu  {\bf 1}_{2\widetilde{N}_c''}  \\
0
\end{array}
\right), 
\qquad
<\Phi'> =
 \left(
\begin{array}{cc}
0  & 0  \\
0 & \Phi'_0  {\bf 1}_{(N_c'-N_c-\widetilde{N}_c'') \otimes i \sigma_2} 
\end{array}
\right) 
\label{poi-1}
\eea
where the zero of $<g>$ is a $
2(N_c'-N_c-\widetilde{N}_c'') \times 2\widetilde{N}_c''$ 
matrix and 
the zeros of $<\Phi'>$ are $2\widetilde{N}_c'' \times 2\widetilde{N}_c''$,
$2\widetilde{N}_c'' \times 2(N_c'-N_c-\widetilde{N}_c'')$, 
and $2(N_c'-N_c-\widetilde{N}_c'') \times
2\widetilde{N}_c''$ matrices.
Then one can expand these fields around on a point (\ref{poi-1}), as
in \cite{ISS} and one arrives at the relevant superpotential
up to quadratic order in the fluctuation. 
At one loop, the effective potential $V_{eff}^{(1)}$ for $\Phi'_0$
leads to the positive value for $m_{\Phi'_0}^2$ implying that these
vacua are stable.

\subsection{Other magnetic theories-III}

By changing the charges of O4-plane in previous brane configuration of
Figure 6A, 
the type IIA brane configuration is realized by 
an ${\cal N}=1$ supersymmetric gauge theory with
\bea
SO(2N_c) \times Sp(N_c') \times SO(2N_c'')
\nonu
\eea
and corresponding matter contents. Then by deforming the theory by 
mass term and taking the magnetic dual on
each gauge group factor, one gets meta-stable brane configurations.
There exists an ${\cal N}=1$ magnetic supersymmetric gauge theory with 
$SO(2\widetilde{N}_c=2N_c'-2N_c+4 ) \times Sp(N_c') \times SO(2N_c'')$ 
with matters which corresponds to Figure 7 with 
opposite O4-plane charges.
Also 
there is an ${\cal N}=1$ magnetic supersymmetric gauge theory with 
$SO(2N_c ) \times Sp(\widetilde{N}_c'=N_c+N_c''-N_c'-2) \times SO(2N_c'')$ 
with matters which corresponds to Figure 9 with 
opposite O4-plane charges. Finally, 
there exists an ${\cal N}=1$ magnetic supersymmetric gauge theory with 
$SO(2N_c ) \times Sp(N_c') \times SO(2\widetilde{N}_c''=2N_c'-2N_c''+4)$ 
with matters which corresponds to Figure 10 with 
opposite O4-plane charges.
The remaining analysis can be done easily
without any difficulty.

\section{Meta-stable brane configurations with six NS-branes plus
  O6-plane}

In this section, we add an orientifold 6-plane to the previous brane
configuration for the product gauge group \cite{BH} realized by three
NS-branes, together with the extra mirrors for them, and 
find out new meta-stable brane configurations.
Or one can realize these brane configurations by inserting the two 
outer NS-branes into the brane configuration \cite{LO,Ahn07-3}.

\subsection{Electric theory}

The type IIA brane configuration  corresponding to 
${\cal N}=1$ supersymmetric gauge theory with
gauge group
\bea
Sp(N_c) \times SU(N_c') \times SU(N_c'') 
\label{gag}
\eea
and with a field $F$ charged under
$({\bf 2N_c, \overline{N_c'}})$, a field $G$ charged under
$({\bf N_c', \overline{N_c''}})$, and their conjugates 
$\widetilde{F}$ and $\widetilde{G}$ 
can be described by 
the left $NS5_L'$-brane, 
the  
NS5-brane, the right $NS5_R'$-brane(and their mirrors),
 $2N_c$-, $N_c'$-  and $N_c''$-color D4-branes as well as
 O6-plane(0123789)
\footnote{From now on, when we say about NS-branes(NS5-brane or
  NS5'-brane), 
they refer to those in positive region of $x^6$. Their mirrors in the
negative region of $x^6$ are understood with O6-plane while we are
taking the brane motion. In other words, there exist three NS-branes:
$NS5_L'$-brane,
NS5-brane and $NS5_R'$-brane from Figure 11A.}. 
The $O6^{-}$-plane acts as $(x^4,x^5,x^6) \rightarrow
(-x^4,-x^5,-x^6)$ and has RR charge $-4$.

Let us place an O6-plane at the origin $x^6=0$
and let us denote the $x^6$ 
coordinates for the $NS5_L'$-brane, the NS5-brane and the $NS5_R'$-brane 
by $x^6=y_1, y_1+y_2, y_1+y_2+y_3$
respectively. Their mirrors can be understood similarly.
The $2N_c$ D4-branes 
are suspended between the 
$NS5_L'$-brane and its mirror, 
the $N_c'$ D4-branes 
are suspended between the 
$NS5_L'$-brane and the NS5-brane(and their mirrors), and 
the $N_c''$ D4-branes  
are suspended between the NS5-brane and the $NS5_R'$-brane(and their mirrors).
We draw this brane configuration in Figure 11A for the vanishing mass
for the field $G$. 

The gauge couplings of $Sp(N_c)$, $SU(N_c')$ and $ SU(N_c'')$
are given by a string coupling constant $g_s$, a string scale $\ell_s$ 
and the $x^6$ coordinates $y_i$ for three NS-branes through
\bea
g_1^2 =\frac{g_s \ell_s}{2y_1}, \qquad 
g_2^2 = \frac{g_s \ell_s}{y_2}, \qquad
g_3^2=\frac{g_s \ell_s}{y_3}.
\nonu
\eea
As $y_3$ goes to $\infty$, the  
$SU(N_c'')$ gauge group becomes a
global symmetry and the theory leads to SQCD with the gauge group
$Sp(N_c) \times SU(N_c')$ and $N_c''$ flavors 
in the fundamental representation.

There is no superpotential in Figure 11A. Let us deform this theory.
Displacing the two NS5'-branes relative each other in the $+v$
direction corresponds to turning on a quadratic
mass-deformed superpotential
for the field $G$ as follows:
\bea
W = m G \widetilde{G} \equiv m \Phi''
\label{Mass}
\eea
where 
the second gauge group indices in $G$ and $\widetilde{G}$ 
are contracted and the mass $m$ is given by (\ref{m}).
The gauge-singlet $\Phi''$ for the second dual gauge group is in the 
adjoint representation for the third dual gauge group, 
i.e., ${(\bf 1, 1, {N_c''}^2-1)  \oplus (1,1,1)}$ 
under the dual gauge group (\ref{dualnew}). 
The $\Phi''$ is a $N_c'' \times N_c''$ matrix.
The $NS5_R'$-brane together with $N_c''$-color D4-branes 
is moving to the $+v$ direction  for
fixed other branes during this mass deformation(and their mirrors to
$-v$ direction). 
Then the $x^5$ coordinate 
of $NS5_L'$-brane is equal to
zero
while the $x^5$ coordinate of $NS5_R'$-branes is given by 
$ \Delta x$.
Giving an expectation value to the meson field $\Phi''$
corresponds to recombination of $N_c'$- and $N_c''$- color 
D4-branes, which will become $N_c''$ or $N_c'$-color D4-branes
in Figure 11A such that they are suspended between 
the $NS5_L'$-brane and the $NS5_R'$-brane 
and pushing them into the $w$
direction. We assume that the number of colors satisfies
\bea
2N_c+N_c'' \geq N_c' \geq 2N_c.
\nonu
\eea

Now 
we draw this brane configuration in Figure 11B for nonvanishing mass
for the fields $G$ and $\widetilde{G}$. 
The geometry for three NS-branes in Figure 11B 
is the same as the one given by first
three NS-branes in Figure 1B.

\begin{figure}[ht]
   \epsfxsize=5.0in 
\centerline{\epsffile{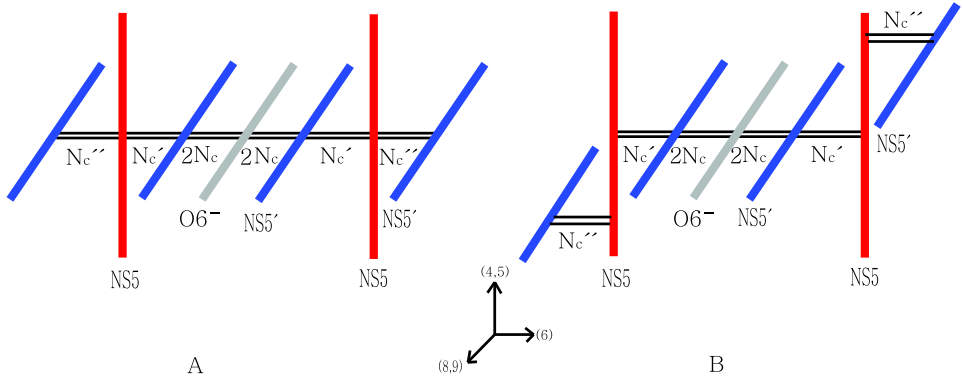}}
   \caption[FIG. \arabic{figure}.]{ 
The  
  ${\cal N}=1$ supersymmetric 
electric brane configuration for the gauge group $Sp(N_c) \times
SU(N_c') \times SU(N_c'')$ 
and  bifundamental $F,\widetilde{F}, G$ and $\widetilde{G}$  
with vanishing(11A) and 
nonvanishing(11B) mass
for the bifundamental $G$ and $\widetilde{G}$. 
In 11B, the $NS5_R'$-brane together with $N_c''$ D4-branes 
is moving to $+v$ direction(and their mirrors to $-v$ direction).  }
\end{figure}

\subsection{Magnetic theory}

By applying the Seiberg dual to the $SU(N_c')$ factor in 
(\ref{gag}), the $NS5_{L,R}'$-branes can be located at the
outside of the two NS5-branes, as in Figure 12.
Starting from Figure 11B and 
interchanging the $NS5_L'$-brane and the NS5-brane(and their mirrors),
one obtains the Figure 12A.

\begin{figure}[ht]
   \epsfxsize=5.0in 
\centerline{\epsffile{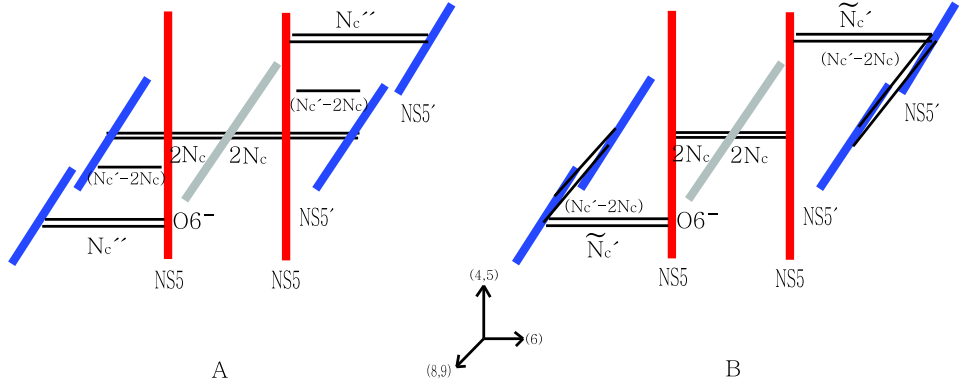}}
   \caption[FIG. \arabic{figure}.]{ 
The ${\cal N}=1$ magnetic brane configuration for the gauge group
$Sp(N_c) \times SU(\widetilde{N}_c'=2N_c+N_c''-N_c') \times SU(N_c'')$ 
corresponding to Figure 11B with D4-
and $\overline{D4}$-branes(12A) and with 
a misalignment between D4-branes(12B) when the NS5'-branes are close to
each other. 
 The number of tilted D4-branes is equal to $N_c'-2N_c=
N_c''-\widetilde{N}_c'$ in 12B. 
The notation for the anti D4-branes is used for the bar
on the number of those branes in 12A.  }
\end{figure}

Before arriving at the Figure 12A, there exists an intermediate 
step where 
the $(N_c''-N_c'+2N_c)$ D4-branes are connecting between the 
NS5-brane and the  $NS5_L'$-brane,  
$N_c''$ D4-branes are connecting between the  $NS5_L'$-brane and   
$NS5_R'$-brane(and their mirrors) as well as $2N_c$ D4-branes between
the
NS5-brane and its mirror.
By reconnecting the $N_c''$ D4-branes 
connecting between the 
NS5-brane and the  $NS5_L'$-brane
with  
the $N_c''$ D4-branes connecting between  
$NS5_L'$-brane 
and the $NS5_R'$-brane and moving those combined
$N_c''$ 
D4-branes
to $+v$-direction(and their mirrors to $-v$ direction), 
one gets the final Figure 12A where we are left with 
$(N_c'-2N_c)$ 
anti-D4-branes between the NS5-brane and   
$NS5_L'$-brane.
When two NS5'-branes in Figure 12A are close to each other, it becomes 
Figure 12B
by realizing that the number of $N_c''$
D4-branes connecting between NS5-brane and $NS5_R'$-brane in Figure
12A can
be rewritten as $(N_c'-2N_c)$ plus $\widetilde{N}_c'$. 
The brane configuration consisting of NS5-brane and two NS5'-branes in
Figure 12B is
exactly the same as those in Figure 2B.  

The dual gauge group is given by 
\bea
Sp(N_c) \times SU(\widetilde{N}_c'=2N_c+N_c''-N_c') \times SU(N_c''). 
\label{dualnew}
\eea
The matter contents are the field $f$ 
 charged under
$({\bf 2N_c, \overline{\widetilde{N}_c'}, 1})$, a field $g$ charged under
$({\bf 1, \widetilde{N}_c', \overline{N_c''}})$, and their conjugates 
$\widetilde{f}$ and $\widetilde{g}$ under the dual gauge group
(\ref{dualnew})
and  
the gauge-singlet $\Phi''$ for the second dual gauge group in the 
adjoint representation for the third dual gauge group, 
i.e.,  ${(\bf 1, 1, {N_c''}^2-1)  \oplus (1,1,1)}$ under the 
dual gauge group.
Then the  $\Phi''$ is a $N_c'' \times N_c''$ matrix.

The cubic superpotential with the mass term (\ref{Mass}) 
is given by
\bea
W_{dual} = \Phi'' g \widetilde{g} + m \Phi''. 
\label{superpo1new}
\eea
Here the magnetic fields $g$ and $\widetilde{g}$  
correspond to 4-4 strings connecting 
the $\widetilde{N}_c'$-color D4-branes(that are 
connecting between the NS5-brane
and the $NS5_R'$-brane in Figure 12B) with $N_c''$-flavor 
D4-branes(which  are realized 
as corresponding D4-branes in Figure 12A).
Although the $N_c''$ D4-branes in Figure 12A cannot move any
directions,
the tilted $(N_c'-2N_c)$-flavor D4-branes can move 
$w$ direction in Figure 12B(and its mirrors).
The remaining upper $\widetilde{N}_c'$ D4-branes are fixed also and cannot 
move any direction. 
Note that 
there is a decomposition 
\bea
N_c''=(N_c'-2N_c)+\widetilde{N}_c'.
\nonu
\eea

The brane configuration for zero mass for the bifundamental,
which has only a cubic superpotential,
can be obtained from Figure 12A by moving
the upper  NS5'-brane(or $NS5_R'$-brane) 
together with $N_c''$ color D4-branes 
into the origin $v=0$(and their mirrors).
Then the number of dual colors for D4-branes 
becomes $2N_c$ between the NS5-brane and its mirror, 
 $\widetilde{N}_c'$ between NS5-brane and $NS5_L'$-brane
and $N_c''$ 
between $NS5_L'$-brane and $NS5_R'$-brane.
Or starting from Figure 11A and moving the 
NS5-brane to the left all the
way past the $NS5_L'$-brane(and their mirrors),
one also obtains the corresponding magnetic brane configuration
for massless case.

The brane configuration in Figure 12A is stable as long as the
distance $\Delta x$ between the upper NS5'-brane and 
the lower NS5'-brane is large. 
If they are close to each other, then this brane
configuration is unstable to decay and leads to 
the brane configuration in Figure
12B.
One can regard these brane configurations as particular states in the
magnetic gauge theory with the gauge group (\ref{dualnew}) and
superpotential (\ref{superpo1new}).

One can perform similar analysis in our brane configuration 
since one can take into account the behavior of
parameters geometrically in the presence of O6-plane.
Then the  upper $(N_c''-\widetilde{N}_c')$ flavor D4-branes of 
straight brane configuration
of
Figure 12B can bend due to the fact that there exists an attractive
gravitational interaction
between those flavor D4-branes and NS5-brane from the DBI action, by
following the procedure of \cite{GK}, as long as $y_1$ is very large.
Then the mirror of NS5-brane does not affect the flavor D4-branes.
On the other hand, 
if $y_1$ goes to zero, then the mirror of NS5-brane plays the role
of enhancing the strength for the  
NS5-branes and will affect both the energy of bending curve, 
$E_{curved}$,  and $\Delta x$. 
Of course, their mirrors, the lower 
$(N_c''-\widetilde{N}_c')$ flavor D4-branes of 
straight brane configuration
of
Figure 12B can bend and their trajectories connecting 
two NS5'-branes should be preserved under the O6-plane, i.e., ${\bf
Z}_2$ symmetric way.

The low energy dynamics of the magnetic brane configuration 
can be described by the ${\cal N}=1$ supersymmetric gauge theory
with gauge group (\ref{dualnew})
and the gauge couplings for the three gauge group factors are
given by
\bea
g_{1,mag}^2  = \frac{g_s \ell_s}{2(y_1+y_2)}, \qquad 
g_{2,mag}^2 = \frac{g_s \ell_s}{y_2}, \qquad
g_{3,mag}^2  = \frac{g_s \ell_s}{(y_3-y_2)}.
\nonu
\eea

The dual gauge theory has  an adjoint $\Phi''$ of $SU(N_c'')$ and 
bifundamentals $f, \widetilde{f}, g$ and $\widetilde{g}$ under the dual gauge
group (\ref{dualnew}) and the superpotential 
corresponding to Figures 12A and 12B is given by 
\bea
W_{dual} = h \Phi'' g \widetilde{g} - h \mu^2 \Phi'', \qquad h^2 = g_{3,
  mag}^2,
\qquad \mu^2 = -\frac{\Delta x}{ 2\pi g_s \ell_s^3}.
\nonu
\eea
Then $ g \widetilde{g}$ is a $\widetilde{N}_c' \times \widetilde{N}_c'$ 
matrix where the third gauge group indices for $g$ and $\widetilde{g}$ 
are contracted with those
of $\Phi''$ while $\mu^2$ is a 
$N_c'' \times N_c''$ matrix.
The product $g \widetilde{g}$ has the same representation for the 
product of quarks
and moreover, 
the third gauge group indices for the field $\Phi''$ play the
role of the flavor indices.

When the upper NS5'-brane(or $NS5_R'$-brane) 
is replaced by coincident $N_c''$ 
D6-branes in Figure 12B, this brane configuration looks similar to the one 
found in \cite{Ahn07-3} where the gauge group was given by 
$SU(n_f+2n_c'-n_c) \times Sp(n_c')$ 
with $n_f$ multiplets  and singlets. 
Then the present $2N_c$ corresponds to the $2n_c'$,
$N_c'$ corresponds to $n_c$,
and 
$N_c''$ corresponds to the $n_f$ of \cite{Ahn07-3}. 

Therefore, the F-term equation, the derivative $W_{dual}$ with respect to the
meson field $\Phi''$ cannot be satisfied if the $N_c''$ exceeds
$\widetilde{N}_c'$.
So the supersymmetry is broken.   
That is, 
there exist three equations from F-term conditions:
$
g\widetilde{g} -\mu^2 =0$ and $ \Phi'' g =0=\widetilde{g} \Phi''$.
Then the solutions for these
are given by 
\bea
<g>   = 
\left(
\begin{array}{c}
\mu  {\bf 1}_{\widetilde{N}_c'}  \\
0
\end{array}
\right), 
\qquad
<\widetilde{g}>   = 
\left(
\begin{array}{cc}
\mu  {\bf 1}_{\widetilde{N}_c'} & 0  \\
\end{array}
\right), 
\qquad
<\Phi''> =
 \left(
\begin{array}{cc}
0  & 0  \\
0 & \Phi_0''  {\bf 1}_{(N_c''-\widetilde{N}_c')} 
\end{array}
\right) 
\label{point20}
\eea
where the zero of $<g>$ is a $
(N_c''-\widetilde{N}_c') \times \widetilde{N}_c'$ 
matrix, the zero of $<\widetilde{g}>$ is a
$\widetilde{N}_c' \times (N_c''-\widetilde{N}_c') $ matrix and 
the zeros of $<\Phi''>$ are $\widetilde{N}_c' \times \widetilde{N}_c'$,
$\widetilde{N}_c' \times 
(N_c''-\widetilde{N}_c')$ and $(N_c''-\widetilde{N}_c') \times
\widetilde{N}_c'$ 
matrices.
Then one can expand these fields around on a point (\ref{point20}), as
in \cite{ISS} and one arrives at the relevant superpotential
up to quadratic order in the fluctuation. 
At one loop, the effective potential $V_{eff}^{(1)}$ for $\Phi_0''$
leads to the positive value for $m_{\Phi_0''}^2$ implying that these
vacua are stable.

\subsection{Other magnetic theory }

Let us consider other magnetic theory for the same electric theory
given in the subsection 4.1.
By applying the Seiberg dual to the $SU(N_c'')$ factor in 
(\ref{gag}), the $NS5_{L,R}'$-branes can be located at the
inside of the two NS5-branes, as in Figure 14.
Starting from Figure 13B and 
interchanging the NS5-brane and the $NS5_R'$-brane(and their mirrors),
one obtains the Figure 14A.
The geometry for three NS-branes in Figure 13B 
is the same as the one given by first
three NS-branes in Figure 3B.

\begin{figure}[ht]
   \epsfxsize=5.0in 
\centerline{\epsffile{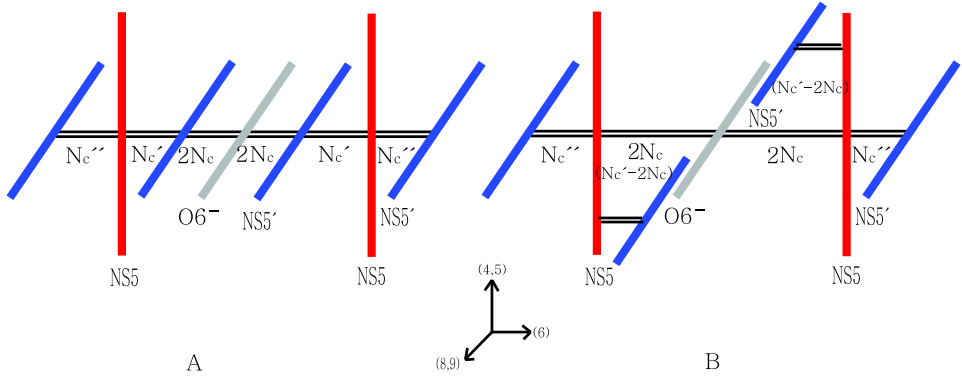}}
   \caption[FIG. \arabic{figure}.]{ 
The  
  ${\cal N}=1$ supersymmetric 
electric brane configuration for the gauge group $Sp(N_c) \times
SU(N_c') \times SU(N_c'')$ 
and the bifundamentals  with vanishing(13A) which is the same as the 
Figure 11A and 
nonvanishing(13B) mass
for the bifundamental $G$ and $\widetilde{G}$. 
This deformation is different from the one 
in (\ref{Mass}).
The $N_c'$ D4-branes in 13A are decomposed into 
$(N_c'-2N_c)$ D4-branes which are moving to $+v$ direction in 13B 
and $2N_c$ D4-branes which are recombined with those D4-branes
connecting between $NS5_L'$-brane and its mirror in 13B.  }
\end{figure}

Before arriving at the Figure 14A, there exists an intermediate 
step where 
the $(N_c'-2N_c)$ D4-branes are connecting between the 
$NS5_L'$-brane and the  $NS5_R'$-brane,  
$(N_c'-N_c'')$ D4-branes are connecting between the  $NS5_R'$-brane and   
NS5-brane(and their mirrors) as well as $2N_c$ D4-branes between
$NS5_R'$-brane and its mirror.
By reconnecting the $(N_c'-2N_c)$ D4-branes 
connecting between the 
$NS5_L'$-brane and the  $NS5_R'$-brane
with  
the $(N_c'-2N_c)$ D4-branes connecting between  
$NS5_R'$-brane 
and the NS5-brane where we introduce $-2N_c$ D4-branes and
$-2N_c$ anti D4-branes and moving those combined 
D4-branes
to $+v$-direction(and their mirrors to $-v$ direction), 
one gets the final Figure 14A where we are left with 
$(N_c''-2N_c)$ 
anti-D4-branes between the $NS5_R'$-brane and   
the NS5-brane.
We assume  that the number of colors satisfies
\bea
N_c' \geq N_c'' \geq 2N_c.
\nonu
\eea
When two NS5'-branes in Figure 14A are close to each other, then 
it leads to Figure 14B
by realizing that the number of $(N_c'-2N_c)$
D4-branes connecting between $NS5_L'$-brane and NS5-brane in Figure
14A can
be rewritten as $(N_c''-2N_c)$ plus $\widetilde{N}_c''$.
The brane configuration consisting of NS5-brane and two NS5'-branes in
Figure 14B is
exactly the same as those in Figure 4B. 

\begin{figure}[ht]
   \epsfxsize=5.0in 
\centerline{\epsffile{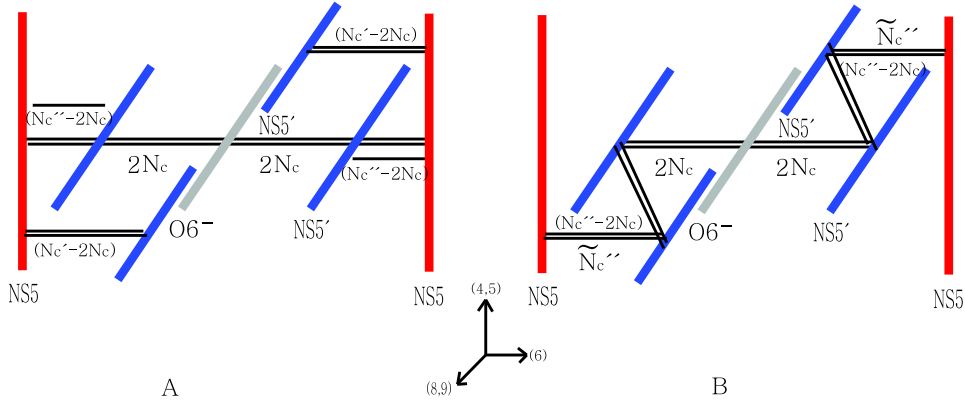}}
   \caption[FIG. \arabic{figure}.]{ 
The  ${\cal N}=1$ magnetic brane configuration for the gauge group
$Sp(N_c) \times SU(N_c') \times SU(\widetilde{N}_c''=N_c'-N_c'')$ 
corresponding to Figure 13B with D4-
and $\overline{D4}$-branes(14A) and with 
a misalignment between D4-branes(14B) when the NS5'-branes are close to
each other.
The number of tilted D4-branes in 14B can be written as
$N_c''-2N_c=(N_c'-2N_c)-\widetilde{N}_c''$.  }
\end{figure}

The dual gauge group is given by
\bea
Sp(N_c) \times SU(N_c') \times SU(\widetilde{N}_c''=N_c'-N_c'') 
\label{dualneww}
\eea
The matter contents are the field $f$ 
 charged under
$({\bf 2N_c, \overline{N_c'}, 1})$, a field $g$ charged under
$({\bf 1, N_c', \overline{\widetilde{N}_c''}})$ 
 and their conjugates 
$\widetilde{f}$ and $\widetilde{g}$
under the dual gauge group
(\ref{dualneww})
and  
the gauge-singlet $\Phi'$ which is in the 
adjoint representation for the second dual gauge group, 
in other words,   
$ ({ \bf   1,  (N_c'-2N_c)^2-1,1})  \oplus  ({\bf 1,1,1})$ under the 
dual gauge group (\ref{dualneww}).
Then the $\Phi'$ is a $(N_c'-2N_c) \times (N_c'-2N_c)$ matrix.
Only $(N_c'-2N_c)$ D4-branes can participate in the mass deformation.

The cubic superpotential with the mass term
is given by
\bea
W_{dual} = \Phi' g \widetilde{g} + m \Phi'
\label{superpo11}
\eea
where we define $\Phi'$ as $\Phi' \equiv G \widetilde{G}$ and 
the third gauge group indices in $G$ and $\widetilde{G}$ 
are contracted, each second gauge group index in them is encoded in 
$\Phi'$. Although the $\Phi'$ that has second gauge group indices 
looks similar to the previous 
$\Phi''$ that has third gauge group indices, 
the group indices are different. 
Here the magnetic fields $g$ and $\widetilde{g}$  
correspond to 4-4 strings connecting 
the $\widetilde{N}_c''$-color D4-branes(that are 
connecting between the $NS5_L'$-brane
and the NS5-brane in Figure 14B) with $N_c'$-flavor 
D4-branes.
Among these $N_c'$-flavor D4-branes, only the strings ending on
the upper $(N_c'-N_c'')$ D4-branes and 
on the tilted  $(N_c''-2N_c)$ 
D4-branes in Figure 14B enter the cubic superpotential term. 
Although the $(N_c'-2N_c)$ D4-branes in Figure 14A cannot move any
directions,
the tilted $(N_c''-2N_c)$-flavor D4-branes can move $w$ direction.
The remaining upper $\widetilde{N}_c''$ D4-branes are fixed also and cannot 
move any direction. 
Note that 
there is a decomposition 
\bea
(N_c'-2N_c)=(N_c''-2N_c)+\widetilde{N}_c''.
\nonu
\eea 

The brane configuration for zero mass for the bifundamental,
which has only a cubic superpotential,
can be obtained from Figure 14A by moving
the upper  NS5'-brane together with $(N_c'-2N_c)$ color D4-branes 
into the origin $v=0$(and their mirrors).
Then the number of dual colors for D4-branes 
becomes $2N_c$ between the $NS5_L'$-brane and its mirror, 
$N_c'$ between the $NS5_L'$-brane and the  $NS5_R'$-brane
and $\widetilde{N}_c''$ 
between $NS5_R'$-brane and NS5-brane.
Or starting from Figure 13A and moving the 
NS5-brane to the right all the
way past the $NS5_R'$-brane(and their mirrors),
one also obtains the corresponding magnetic brane configuration
for massless case.

The brane configuration in Figure 14A is stable as long as the
distance $\Delta x$ between the upper NS5'-brane and 
the lower NS5'-brane is large. 
If they are close to each other, then this brane
configuration is unstable to decay and leads to 
the brane configuration in Figure
14B.
One can regard these brane configurations as particular states in the
magnetic gauge theory with the gauge group (\ref{dualneww}) and
superpotential (\ref{superpo11}).
Then the  upper $(N_c'-2N_c-\widetilde{N}_c'')$ flavor D4-branes of 
straight brane configuration
of
Figure 14B can bend due to the fact that there exists an attractive
gravitational interaction
between those flavor D4-branes and NS5-brane from the DBI action, 
as long as $y_1$ is very large.
Of course, their mirrors, the lower 
$(N_c'-2N_c-\widetilde{N}_c'')$ flavor D4-branes of 
straight brane configuration
of
Figure 14B can bend and their trajectories connecting 
two NS5'-branes should be preserved under the O6-plane, i.e., ${\bf
Z}_2$ symmetric way.

When the upper NS5'-brane(or $NS5_L'$-brane) 
is replaced by coincident $(N_c'-2N_c)$ 
D6-branes in Figure 14B, this brane configuration looks similar to the one 
found in \cite{Ahn07-3} where the gauge group was given by 
$SU(n_f+n_c'-n_c) \times SO(n_c')$ 
with $n_f$ multiplets, bifundamentals,  and singlets. 
Then the present $2N_c$ corresponds to the $n_c'$,
$(N_c'-2N_c)$ corresponds to $n_f$,
and 
$N_c''$ corresponds to the $n_c$ of \cite{Ahn07-3}.
Note that $Sp(N_c)$ corresponds to $SO(n_c')$.
Moreover, there is a  
meta-stable brane configuration for the gauge group given by 
$SU(n_c) \times SU(n_f'+n_c-n_c')$ 
with fundamentals, bifundamentals, an antisymmetric flavor, a
conjugate symmetric flavor, and singlets where there are NS5'-brane,
$O6^{\pm}$-planes,
and eight semi infinite D6-branes at $x^6=0$. 
Then the our $2N_c$ corresponds to the $n_c$,
the number $(N_c'-2N_c)$ corresponds to $n_f'$,
and 
our $N_c''$ corresponds to the $n_c'$ of \cite{Ahn07-4}. 

The low energy dynamics of the magnetic brane configuration 
can be described by the ${\cal N}=1$ supersymmetric gauge theory
with gauge group (\ref{dualneww})
and the gauge couplings for the three gauge group factors are
given by
\bea
g_{1,mag}^2  = \frac{g_s \ell_s}{2y_1}, \qquad 
g_{2,mag}^2 = \frac{g_s \ell_s}{(y_2-y_3)}, \qquad
g_{3,mag}^2  = \frac{g_s \ell_s}{y_3}.
\nonu
\eea
The dual gauge theory has  an adjoint $\Phi'$ of $SU(N_c')$ and 
bifundamentals $f, \widetilde{f}, g$ and 
$\widetilde{g}$ under the dual gauge
group (\ref{dualneww}) and the superpotential 
corresponding to Figures 14A and 14B is given by 
\bea
W_{dual} = h \Phi' g \widetilde{g} - h \mu^2 \Phi', \qquad h^2 = g_{2,
  mag}^2,
\qquad \mu^2 = -\frac{\Delta x}{ 2\pi g_s \ell_s^3}.
\nonu
\eea
Then $ g \widetilde{g}$ is a $\widetilde{N}_c'' \times \widetilde{N}_c''$ 
matrix where the second gauge group indices for $g$ and $\widetilde{g}$ 
are contracted with those
of $\Phi'$ while $\mu^2$ is a 
$(N_c'-2N_c) \times (N_c'-2N_c)$ matrix.
The product $g \widetilde{g}$ has the same representation for the 
product of quarks
and moreover, 
the second gauge group indices for the field $\Phi'$ play the
role of the flavor indices, as above.

Therefore, the F-term equation, the derivative $W_{dual}$ with respect to the
meson field $\Phi'$ cannot be satisfied if the $(N_c'-2N_c)$ exceeds
$\widetilde{N}_c''$.
So the supersymmetry is broken.   
That is, 
there exist three equations from F-term conditions:
$
g\widetilde{g} -\mu^2 =0$ and $ \Phi' g =0=\widetilde{g} \Phi'$.
Then the solutions for these
are given by 
\bea
<g>   = 
\left(
\begin{array}{c}
\mu  {\bf 1}_{\widetilde{N}_c''}  \\
0
\end{array}
\right), 
\qquad
<\widetilde{g}>   = 
\left(
\begin{array}{cc}
\mu  {\bf 1}_{\widetilde{N}_c''} & 0  \\
\end{array}
\right), 
\qquad
<\Phi'> =
 \left(
\begin{array}{cc}
0  & 0  \\
0 & \Phi_0'  {\bf 1}_{(N_c'-2N_c)-\widetilde{N}_c''} 
\end{array}
\right) 
\label{point12-1}
\eea
where the zero of $<g>$ is a $
(N_c'-2N_c-\widetilde{N}_c'') \times \widetilde{N}_c''$ 
matrix, the zero of $<\widetilde{g}>$ is a
$\widetilde{N}_c'' \times (N_c'-2N_c-\widetilde{N}_c'') $ matrix and 
the zeros of $<\Phi'>$ are $\widetilde{N}_c'' \times \widetilde{N}_c''$,
$\widetilde{N}_c'' \times 
(N_c'-2N_c-\widetilde{N}_c'')$ and $(N_c'-2N_c-\widetilde{N}_c'') \times
\widetilde{N}_c''$ 
matrices.
Then one can expand these fields around on a point (\ref{point12-1}), as
in \cite{ISS} and one arrives at the relevant superpotential
up to quadratic order in the fluctuation. 
At one loop, the effective potential $V_{eff}^{(1)}$ for $\Phi'_0$
leads to the positive value for $m_{\Phi'_0}^2$ implying that these
vacua are stable.


\subsection{Other magnetic theories }

In this subsection, we add an orientifold 6-plane with positive charge
to the previous brane
configuration for the product gauge group \cite{BH} realized by three
NS-branes, together with the extra mirrors for them, and 
find out new meta-stable brane configurations.
Or one can realize these brane configurations by inserting the two 
outer NS-branes into the brane configuration \cite{LO,Ahn07-3}.

\subsubsection{Electric theory}

The type IIA brane configuration  corresponding to 
${\cal N}=1$ supersymmetric gauge theory with
gauge group
\bea
SO(N_c) \times SU(N_c') \times SU(N_c'') 
\label{gag1}
\eea
and with a field $F$ charged under
$({\bf N_c, \overline{N_c'}})$, a field $G$ charged under
$({\bf N_c', \overline{N_c''}})$, and their conjugates 
$\widetilde{F}$ and $\widetilde{G}$ 
can be described by 
the left $NS5_L$-brane, 
the  
NS5'-brane, the right $NS5_R$-brane(and their mirrors),
 $N_c$-, $N_c'$-  and $N_c''$-color D4-branes as well as
$O6^{+}$-plane(0123789).
The $O6^{+}$-plane acts as $(x^4,x^5,x^6) \rightarrow
(-x^4,-x^5,-x^6)$ and has RR charge $+4$.

Let us place an $O6^{+}$-plane at the origin $x^6=0$
and let us denote the $x^6$ 
coordinates for the $NS5_L$-brane, the NS5'-brane and the $NS5_R$-brane 
by $x^6=y_1, y_1+y_2, y_1+y_2+y_3$
respectively. Their mirrors can be understood similarly.
The $N_c$ D4-branes 
are suspended between the 
$NS5_L$-brane and its mirror, 
the $N_c'$ D4-branes 
are suspended between the 
$NS5_L$-brane and the NS5'-brane(and their mirrors), and 
the $N_c''$ D4-branes  
are suspended between the NS5'-brane and the $NS5_R$-brane(and their mirrors).
We assume that the number of colors satisfies
\bea
N_c+N_c'' \geq N_c' \geq N_c.
\nonu
\eea

\subsubsection{Magnetic theory}

By applying the Seiberg dual to the $SU(N_c')$ factor in 
(\ref{gag1})  and 
interchanging the $NS5_L$-brane and the NS5'-brane(and their mirrors),
one obtains the Figure 15A.

\begin{figure}[ht]
   \epsfxsize=5.0in 
\centerline{\epsffile{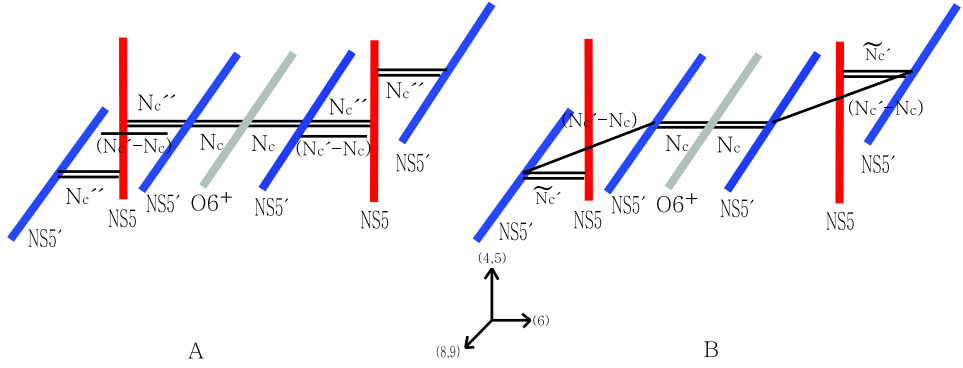}}
   \caption[FIG. \arabic{figure}.]{ 
The ${\cal N}=1$ magnetic brane configuration for the gauge group
$SO(N_c) \times SU(\widetilde{N}_c'=N_c+N_c''-N_c') \times SU(N_c'')$ 
with D4-
and $\overline{D4}$-branes(15A) and with 
a misalignment between D4-branes(15B) when the NS5'-branes are close to
each other. Note that the number of D4-branes on the gauge group 
$SO(N_c)$ is equal to $N_c$ not $2N_c$.
 The number of tilted D4-branes is equal to $N_c'-N_c=
N_c''-\widetilde{N}_c'$ in 15B.
The deformation is rleated to the bifundamentals $G$ and 
$\widetilde{G}$.   }
\end{figure}

Before arriving at the Figure 15A, there exists an intermediate 
step where 
the $(N_c''-N_c'+N_c)$ D4-branes are connecting between the 
NS5'-brane and the  $NS5_L$-brane,  
$N_c''$ D4-branes are connecting between the  $NS5_L$-brane and   
$NS5_R$-brane(and their mirrors) as well as $N_c$ D4-branes between
NS5'-brane and its mirror.
By rotating $NS5_R$-brane by an angle $\frac{\pi}{2}$, moving it with 
$N_c''$ 
D4-branes
to $+v$-direction(and their mirrors to $-v$ direction), 
one gets the final Figure 15A where we are left with 
$(N_c'-N_c)$ 
anti-D4-branes between the $NS5_L'$-brane and   
NS5-brane.
When two NS5'-branes in Figure 15A are close to each other, it becomes 
Figure 15B
 by realizing that the number of $N_c''$
D4-branes connecting between NS5-brane and $NS5_R'$-brane can
be rewritten as $(N_c'-N_c)$ plus $\widetilde{N}_c'$. 

The dual gauge group is given by 
\bea
SO(N_c) \times SU(\widetilde{N}_c'=N_c+N_c''-N_c') \times SU(N_c''). 
\label{dualnew11}
\eea
The matter contents are the field $f$ 
 charged under
$({\bf N_c, \overline{\widetilde{N}_c'}, 1})$, a field $g$ charged under
$({\bf 1, \widetilde{N}_c', \overline{N_c''}})$, and their conjugates 
$\widetilde{f}$ and $\widetilde{g}$ under the dual gauge group
(\ref{dualnew11})
and  
the gauge-singlet $\Phi''$ for the second dual gauge group in the 
adjoint representation for the third dual gauge group, 
i.e.,  ${(\bf 1, 1, {N_c''}^2-1)  \oplus (1,1,1)}$ under the 
dual gauge group.
Then the  $\Phi''$ is a $N_c'' \times N_c''$ matrix.

The cubic superpotential with the mass term  
is given by
(\ref{superpo1new})
where we define $\Phi''$ as $\Phi'' \equiv G \widetilde{G}$ and 
the second gauge group indices in $G$ and $\widetilde{G}$ 
are contracted, each third gauge group index in them is encoded in 
$\Phi''$. Although the $\Phi''$ that has third gauge group indices 
looks similar to the previous 
$\Phi'$ that has second gauge group indices 
the group indices are different.
Here the magnetic fields $g$ and $\widetilde{g}$  
correspond to 4-4 strings connecting 
the $\widetilde{N}_c'$-color D4-branes(that are 
connecting between the NS5-brane
and the $NS5_R'$-brane in Figure 15B) with $N_c''$-flavor 
D4-branes(which  are realized 
as corresponding D4-branes in Figure 15A).
Although the $N_c''$ D4-branes in Figure 15A cannot move any
directions,
the tilted $(N_c'-N_c)$-flavor D4-branes can move 
$w$ direction in Figure 15B(and its mirrors).
The remaining upper $\widetilde{N}_c'$ D4-branes are fixed also and cannot 
move any direction. 
Note that 
there is a decomposition 
\bea
N_c''=(N_c'-N_c)+\widetilde{N}_c'.
\nonu
\eea

The brane configuration for zero mass for the bifundamental,
which has only a cubic superpotential,
can be obtained from Figure 15A by moving
the upper  NS5'-brane together with $N_c''$ color D4-branes 
into the origin $v=0$(and their mirrors).
Then the number of dual colors for D4-branes 
becomes $N_c$ between the $NS5_L'$-brane and its mirror, 
 $\widetilde{N}_c'$ between $NS5_L'$-brane and NS5-brane
and $N_c''$ 
between NS5-brane and $NS5_R'$-brane.

The brane configuration in Figure 15A is stable as long as the
distance $\Delta x$ between the upper NS5'-brane and 
the lower NS5'-brane 
is large. If they are close to each other, then this brane
configuration is unstable to decay to 
the brane configuration in Figure
15B.
One can regard these brane configurations as particular states in the
magnetic gauge theory with the gauge group and
superpotential.
The  upper $(N_c''-\widetilde{N}_c')$ flavor D4-branes of 
straight brane configuration
of
Figure 15B  bend since there exists an attractive
gravitational interaction
between those flavor D4-branes and NS5-brane from the DBI action. 
As mentioned in \cite{Ahn07-5},
the two NS5'-branes are located at different side of NS5-brane in
Figure 15B and the DBI action computation for this bending curve
should be taken into account. 

The low energy dynamics of the magnetic brane configuration 
can be described by the ${\cal N}=1$ supersymmetric gauge theory
with gauge group (\ref{dualnew11})
and the gauge couplings for the three gauge group factors are
given by the expressions in subsection 4.2.

The dual gauge theory has  an adjoint $\Phi''$ of $SU(N_c'')$ and 
bifundamentals $f, \widetilde{f}, g$ and $\widetilde{g}$ under the dual gauge
group (\ref{dualnew11}) and the superpotential 
corresponding to Figures 15A and 15B is given by 
the expressions in subsection 4.2.
Then $ g \widetilde{g}$ is a $\widetilde{N}_c' \times \widetilde{N}_c'$ 
matrix where the third gauge group indices for $g$ and $\widetilde{g}$ 
are contracted with those
of $\Phi''$ while $\mu^2$ is a 
$N_c'' \times N_c''$ matrix.
The product $g \widetilde{g}$ has the same representation for the 
product of quarks
and moreover, 
the third gauge group indices for the field $\Phi''$ play the
role of the flavor indices.

When the upper NS5'-brane(or $NS5_R'$-brane) 
is replaced by coincident $N_c''$ 
D6-branes in Figure 15B, this brane configuration looks similar to the one 
found in \cite{Ahn07-3} where the gauge group was given by 
$SU(n_f+n_c'-n_c) \times Sp(n_c')$ 
with $n_f$ multiplets  and singlets. 
Then the present $N_c$ corresponds to the $n_c'$,
$N_c'$ corresponds to $n_c$,
and 
$N_c''$ corresponds to the $n_f$ of \cite{Ahn07-3}. 

Therefore, the F-term equation, the derivative $W_{dual}$ with respect to the
meson field $\Phi''$ cannot be satisfied if the $N_c''$ exceeds
$\widetilde{N}_c'$.
So the supersymmetry is broken.   
That is, 
there exist three equations from F-term conditions:
$
g\widetilde{g} -\mu^2 =0$ and $ \Phi'' g =0=\widetilde{g} \Phi''$.
Then the solutions for these
are given by 
the expressions in subsection 4.2.
Then one can expand these fields around on a point, as
in \cite{ISS} and one arrives at the relevant superpotential
up to quadratic order in the fluctuation. 
At one loop, the effective potential $V_{eff}^{(1)}$ for $\Phi_0''$
leads to the positive value for $m_{\Phi_0''}^2$ implying that these
vacua are stable.

\subsubsection{Other magnetic theory }

Let us consider other magnetic theory for the same electric theory
given in the subsection 4.4.1.
By applying the Seiberg dual to the $SU(N_c'')$ factor in 
(\ref{gag1})  and 
interchanging the NS5'-brane and the $NS5_R$-brane(and their mirrors),
one obtains the Figure 16A.

Before arriving at the Figure 16A, there exists an intermediate 
step where 
the $N_c'$ D4-branes are connecting between the 
$NS5_L$-brane and the  $NS5_R$-brane,  
$(N_c'-N_c'')$ D4-branes are connecting between the  $NS5_R$-brane and   
NS5'-brane(and their mirrors) as well as $N_c$ D4-branes between
$NS5_L$-brane and its mirror.
By rotating $NS5_L$-brane by an angle $\frac{\pi}{2}$, 
moving it with the $(N_c'-N_c)$ D4-branes 
to $+v$ direction where we introduce $(N_c'-N_c)$ D4-branes and
$(N_c'-N_c)$ anti D4-branes between the $NS5_R$-brane and the 
NS5'-brane(and their mirrors to $-v$ direction), 
one gets the final Figure 16A where we are left with 
$(N_c''-N_c)$ 
anti-D4-branes between the NS5-brane and   
the $NS5_R'$-brane.
We assume that  the number of colors satisfies
\bea
N_c' \geq N_c'' \geq N_c.
\nonu
\eea
When two NS5'-branes in Figure 16A are close to each other, then 
it leads to Figure 16B
 by realizing that the number of $(N_c'-N_c)$
D4-branes connecting between $NS5_L'$-brane and NS5-brane can
be rewritten as $(N_c''-N_c)$ plus $\widetilde{N}_c''$.
The brane configuration consisting of NS5-brane and two NS5'-branes in
Figure 16B is
exactly the same as those in Figure 5B''. 

\begin{figure}[ht]
   \epsfxsize=5.0in 
\centerline{\epsffile{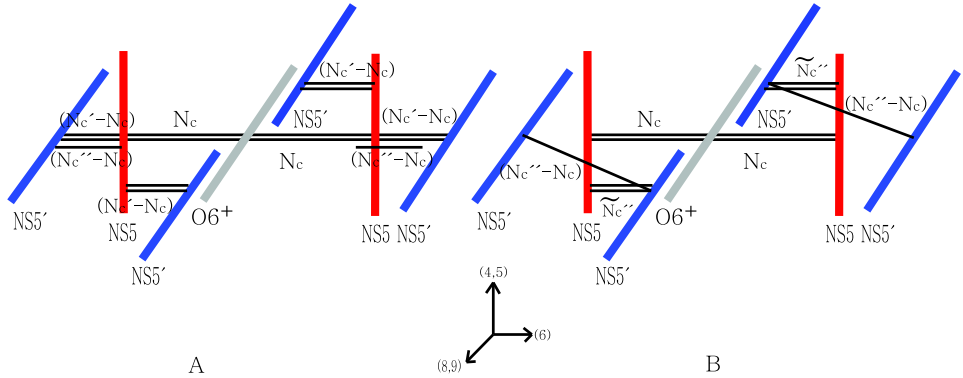}}
   \caption[FIG. \arabic{figure}.]{ 
The  ${\cal N}=1$ magnetic brane configuration for the gauge group
$SO(N_c) \times SU(N_c') \times SU(\widetilde{N}_c''=N_c'-N_c'')$ 
with D4-
and $\overline{D4}$-branes(16A) and with 
a misalignment between D4-branes(16B) when the NS5'-branes are close to
each other.
The number of tilted D4-branes in 16B can be written as
$N_c''-N_c=(N_c'-N_c)-\widetilde{N}_c''$.
The deformation is different from the previous one.  }
\end{figure}

The dual gauge group is given by
\bea
SO(N_c) \times SU(N_c') \times SU(\widetilde{N}_c''=N_c'-N_c'') 
\label{dualneww11}
\eea
The matter contents are the field $f$ 
 charged under
$({\bf N_c, \overline{N_c'}, 1})$, a field $g$ charged under
$({\bf 1, N_c', \overline{\widetilde{N}_c''}})$ 
 and their conjugates 
$\widetilde{f}$ and $\widetilde{g}$
under the dual gauge group
(\ref{dualneww11})
and  
the gauge-singlet $\Phi'$ which is in the 
adjoint representation for the second dual gauge group, 
in other words,   
$ ({ \bf   1,  (N_c'-N_c)^2-1,1})  \oplus  ({\bf 1,1,1})$ under the 
dual gauge group (\ref{dualneww11}).
Then the $\Phi'$ is a $(N_c'-N_c) \times (N_c'-N_c)$ matrix.
Only $(N_c'-N_c)$ D4-branes are participating in the mass deformation.

The cubic superpotential with the mass term
is given by
(\ref{superpo11})
where we define $\Phi'$ as $\Phi' \equiv G \widetilde{G}$ and 
the third gauge group indices in $G$ and $\widetilde{G}$ 
are contracted, each second gauge group index in them is encoded in 
$\Phi'$. Although the $\Phi'$ that has second gauge group indices 
looks similar to the previous 
$\Phi''$ that has third gauge group indices, 
the group indices are different.
Here the magnetic fields $g$ and $\widetilde{g}$  
correspond to 4-4 strings connecting 
the $\widetilde{N}_c''$-color D4-branes(that are 
connecting between the $NS5_L'$-brane
and the NS5-brane in Figure 16B) with $N_c'$-flavor 
D4-branes.
Among these $N_c'$-flavor D4-branes, only the strings ending on
the upper $(N_c'-N_c'')$ D4-branes and 
on the tilted  $(N_c''-N_c)$ 
D4-branes in Figure 16B enter the cubic superpotential term. 
Although the $(N_c'-N_c)$ D4-branes in Figure 16A cannot move any
directions,
the tilted $(N_c''-N_c)$-flavor D4-branes can move $w$ direction.
The remaining upper $\widetilde{N}_c''$ D4-branes are fixed also and cannot 
move any direction. 
Note that 
there is a decomposition 
\bea
(N_c'-N_c)=(N_c''-N_c)+\widetilde{N}_c''.
\nonu
\eea

The brane configuration for zero mass for the bifundamental,
which has only a cubic superpotential,
can be obtained from Figure 16A by moving
the upper  NS5'-brane together with $(N_c'-N_c)$ color D4-branes 
into the origin $v=0$(and their mirrors).
Then the number of dual colors for D4-branes 
becomes $N_c$ between the $NS5_L'$-brane and its mirror, 
$N_c'$ between the $NS5_L'$-brane and the  NS5-brane
and $\widetilde{N}_c''$ 
between NS5-brane and $NS5_R'$-brane.

When the upper NS5'-brane(or $NS5_L'$-brane) 
is replaced by coincident $(N_c'-N_c)$ 
D6-branes in Figure 16B, this brane configuration looks similar to the one 
found in \cite{Ahn07-3} where the gauge group was given by 
$SU(n_f+2n_c'-n_c) \times Sp(n_c')$ 
with $n_f$ multiplets, bifundamentals, and singlets. 
Then the present $N_c$ corresponds to the $2n_c'$,
$(N_c'-N_c)$ corresponds to $n_f$,
and 
$N_c''$ corresponds to the $n_c$ of \cite{Ahn07-3}. 
Note that $SO(N_c)$ corresponds to $Sp(n_c')$.
Moreover, 
the meta-stable brane configuration corresponding to gauge group given by 
$SU(n_c) \times SU(n_f'+n_c-n_c')$ 
with fundamentals, bifundamentals, 
a symmetric flavor, a conjugate symmetric
flavor,  
and singlets was given in \cite{Ahn07-4} where there exists NS5-brane
on the O6-plane. 
Then our $N_c$ corresponds to the $n_c$,
our $(N_c'-N_c)$ corresponds to $n_f'$,
and 
our $N_c''$ corresponds to the $n_c'$.

The brane configuration in Figure 16A is stable as long as the
distance $\Delta x$ between the upper NS5'-brane and 
the lower NS5'-brane 
is large. If they are close to each other, then this brane
configuration is unstable to decay to 
the brane configuration in Figure
16B.
One can regard these brane configurations as particular states in the
magnetic gauge theory with the gauge group and
superpotential.
The  upper $(N_c'-N_c-\widetilde{N}_c'')$ flavor D4-branes of 
straight brane configuration
of
Figure 16B  bend since there exists an attractive
gravitational interaction
between those flavor D4-branes and NS5-brane from the DBI action. 
As mentioned in \cite{Ahn07-5},
the two NS5'-branes are located at different side of NS5-brane in
Figure 16B and the DBI action computation for this bending curve
should be taken into account. 

The low energy dynamics of the magnetic brane configuration 
can be described by the ${\cal N}=1$ supersymmetric gauge theory
with gauge group (\ref{dualneww11})
and the gauge couplings for the three gauge group factors are
given by
the expressions in subsection 4.3.
The dual gauge theory has  an adjoint $\Phi'$ of $SU(N_c')$ and 
bifundamentals $f, \widetilde{f}, g$ and $\widetilde{g}$ under the dual gauge
group (\ref{dualneww11}) and the superpotential 
corresponding to Figures 16A and 16B is given by 
the one in subsection 4.3.
Then $ g \widetilde{g}$ is a $\widetilde{N}_c'' \times \widetilde{N}_c''$ 
matrix where the second gauge group indices for $g$ and $\widetilde{g}$ 
are contracted with those
of $\Phi'$ while $\mu^2$ is a 
$(N_c'-N_c) \times (N_c'-N_c)$ matrix.
The product $g \widetilde{g}$ has the same representation for the 
product of quarks
and moreover, 
the second gauge group indices for the field $\Phi'$ play the
role of the flavor indices.

Therefore, the F-term equation, the derivative $W_{dual}$ with respect to the
meson field $\Phi'$ cannot be satisfied if the $(N_c'-N_c)$ exceeds
$\widetilde{N}_c''$.
So the supersymmetry is broken.   
That is, 
there exist three equations from F-term conditions:
$
g\widetilde{g} -\mu^2 =0$ and $ \Phi' g =0=\widetilde{g} \Phi'$.
Then the solutions for these
are given by 
\bea
<g>   = 
\left(
\begin{array}{c}
\mu  {\bf 1}_{\widetilde{N}_c''}  \\
0
\end{array}
\right), 
\qquad
<\widetilde{g}>   = 
\left(
\begin{array}{cc}
\mu  {\bf 1}_{\widetilde{N}_c''} & 0  \\
\end{array}
\right), 
\qquad
<\Phi'> =
 \left(
\begin{array}{cc}
0  & 0  \\
0 & \Phi_0'  {\bf 1}_{(N_c'-N_c)-\widetilde{N}_c''} 
\end{array}
\right) 
\label{point12}
\eea
where the zero of $<g>$ is a $
(N_c'-N_c-\widetilde{N}_c'') \times \widetilde{N}_c''$ 
matrix, the zero of $<\widetilde{g}>$ is a
$\widetilde{N}_c'' \times (N_c'-N_c-\widetilde{N}_c'') $ matrix and 
the zeros of $<\Phi'>$ are $\widetilde{N}_c'' \times \widetilde{N}_c''$,
$\widetilde{N}_c'' \times 
(N_c'-N_c-\widetilde{N}_c'')$ and $(N_c'-N_c-\widetilde{N}_c'') \times
\widetilde{N}_c''$ 
matrices.
Then one can expand these fields around on a point (\ref{point12}), as
in \cite{ISS} and one arrives at the relevant superpotential
up to quadratic order in the fluctuation. 
At one loop, the effective potential $V_{eff}^{(1)}$ for $\Phi'_0$
leads to the positive value for $m_{\Phi'_0}^2$ implying that these
vacua are stable.

%

\section{Conclusions and outlook}

The meta-stable brane configurations we have found are summarized by
Figures 2, 4, 5, 7, 9, 10, 12, 14, 15 and 16.
If we replace the NS5'-brane in Figures 2B, 7B with
opposite O4-plane charge, 14B with opposite O6-plane charge, 
and 16B with opposite O6-plane charge, with the coincident D6-branes, 
those brane configurations become nonsupersymmetic
minimal energy brane configurations found in \cite{Ahn07-3}, 
in \cite{Ahn07-2}, in \cite{Ahn07-3}, and in \cite{Ahn07-3}, respectively.

So far, we have 
considered the cases for even number of NS-branes, i.e., four and six.
For odd cases, i.e., three and five NS-branes, 
the construction of meta-stable brane configuration 
has been done in \cite{Ahn07-5}.
So it is natural to ask what happens if 
there are seven NS-branes.
When this extra seventh NS-brane is located at the O6-plane in section
4,
then the gauge group will be the same as the one in section 2, i.e., 
$
SU(N_c) \times SU(N_c') \times SU(N_c'')$ with different matter contents.
This can be obtained also 
from the brane configuration of \cite{Ahn07-4} by
adding two outer NS-branes. It would be interesting to find out how
the meta-stable brane configurations appear.

Some different directions on the meta-stable vacua
are present in
recent relevant works \cite{MPS}-\cite{ABSV} where 
some of them are described in the type IIB string theory.
It would be very interesting to find out
how the meta-stable brane configurations from 
type IIA string theory including the present work are related to
those brane configurations from type IIB string theory.

\vspace{.7cm}

\centerline{\bf Acknowledgments}

I would like to thank 
D. Kutasov 
for discussions. 
I would like to thank Kyungho Oh, who passed away from cancer, for
ongoing collaboration and discussions during the last 10 years and,
in memory of him, 
I would like to dedicate 
this work to him. 
This work was supported by grant No.
R01-2006-000-10965-0 from the Basic Research Program of the Korea
Science \& Engineering Foundation.

\end{document}